\documentclass[preprint]{aastex}

\slugcomment{}
\usepackage{bm}
\usepackage{graphicx}
\usepackage{lscape}

\shorttitle{Three-dimensional MHD Simulations of Jets from Accretion Disks}
\shortauthors{Kigure & Shibata}

\begin{document}

\title{Three-dimensional MHD Simulations of Jets from Accretion Disks}

\author{Hiromitsu Kigure\altaffilmark{1},Kazunari Shibata\altaffilmark{1}}
\altaffiltext{1}{Kwasan and Hida Observatories, Kyoto University,
Yamashina, Kyoto 607-8471, Japan;kigure@kwasan.kyoto-u.ac.jp}

\begin{abstract}
We report the results of 3-dimensional magnetohydrodynamic (MHD) simulations 
of a jet formation by the interaction between an accretion disk and a large 
scale magnetic field. The disk is not treated as a boundary condition but 
is solved self-consistently. To investigate the stability of MHD jet, the 
accretion disk is perturbed with a non-axisymmetric sinusoidal or random 
fluctuation of the rotational velocity. 
The dependences of the jet velocity $(v_z)$, mass 
outflow rate $(\dot{M}_w)$, and mass accretion rate $(\dot{M}_a)$ on the 
initial magnetic field strength in both non-axisymmetric cases are 
similar to those in the axisymmetric case. That is, $v_z \propto B_0^{1/3}$, 
$\dot{M}_w \propto B_0$ and $\dot{M}_a \propto B_0^{1.4}$ where $B_0$ is the 
initial magnetic field strength. 
The former two relations are consistent with the Michel's steady solution, 
$v_z \propto \left( B_0^2/\dot{M}_w \right)^{1/3}$, although the jet and 
accretion do not reach the steady state.
In both perturbation cases, a non-axisymmetric structure with $m=2$ 
appears in the jet, where $m$ means the azimuthal wave number. This 
structure can not be explained by Kelvin-Helmholtz instability and seems to 
originate in the accretion disk. 
Non-axisymmetric modes in the jet reach almost constant levels after about 1.5 
orbital periods of the accretion disk, 
while all modes in the accretion disk grow with oscillation. 
As for the angular momentum transport by Maxwell stress, the vertical 
component, $\langle B_\phi B_z/4\pi \rangle$, is comparable to the radial 
component, $\langle B_\phi B_r/4\pi \rangle$, in the wide range of initial 
magnetic field strength.
\end{abstract}


\section{Introduction}
Astrophysical jets are one of the most important and interesting subject in 
astrophysics. They have been observed in various spatial scales, 
e.g. in young stellar objects (YSOs) (e.g., Fukui et al. 1993; Ogura 1995; 
Furuya et al. 1999), active galactic nuclei (e.g., Bridle \& Perley 1984; 
Kubo et al. 1998), and X-ray binaries (e.g., Margon 1984; Mirabel \& 
Rodriguez 1994; Kotani et al. 1996). The acceleration and collimation 
mechanisms of jets are still not made clear and various models have been 
proposed (e.g., Ferrari 1998; Meier, Koide, \& Uchida 2001; Bisnovatyi-Kogan 
\& Lovelace 2001). 

One of the most promising models is 
magnetohydrodynamic (MHD) acceleration from accretion disks. Lovelace (1976) 
and Blandford (1976) first pointed out that electromagnetic force can extract 
energy and angular momentum from accretion disks and that the plasma is 
accelerated relativistically in two opposite directions. Blandford \& Payne 
(1982) showed that a magneto-centrifugally driven outflow from a Keplerian 
disk is possible when an angle between the poloidal magnetic field lines and 
the surface of the accretion disk is less than $60^\circ$. They discussed 
self-similar solutions of the steady and axisymmetric MHD equations and the 
possibility of such acceleration and collimation of the flow from a cold 
Keplerian disk. Kudoh \& Shibata (1995, 1997a) solved 1-dimensional steady and 
axisymmetric MHD equations and investigated properties of steady MHD jet. 
They found that the steady MHD jet can be divided into two classes; one is 
the magneto-centrifugally driven jet when the magnetic field is strong and the 
other is the magnetic pressure driven jet when the field is weak. The former 
has dependences of the jet velocity ($V_{jet}$) and mass outflow rate 
($\dot{M}_w = dM_w/dt$) on magnetic energy ($E_{mg}$) as $V_{jet} \propto 
E_{mg}^{1/3}$, $\dot{M}_w \sim$ constant and the latter has dependences 
as $V_{jet} \propto E_{mg}^{1/6}$, $\dot{M}_w \propto E_{mg}^{0.5}$. 
This mass flux scaling law was confirmed by Ustyugova et al. (1999).

On the other hand, Uchida \& Shibata (1985) and Shibata \& Uchida (1986, 1987) 
performed time-dependent 2-dimensional axisymmetric 
(2.5-dimensional) MHD numerical simulations of magnetically driven jets from 
accretion disks. They solved the interaction 
between a geometrically thin rotating disk and a large-scale magnetic field 
that was initially uniform and vertical. Shibata \& Uchida (1986) investigated 
the dependence of the jet velocity on the magnetic field strength. The jet 
velocity increases as the magnetic field strength increases in a manner 
similar to Michel's scaling law (Michel 1969) but is typically of the order of 
the disk's 
Keplerian velocity. Kudoh, Matsumoto, \& Shibata (1998) showed the following 
results: The dependences of the jet velocity ($v_z$) and mass outflow rate 
($\dot{M}_w$) on the initial magnetic field strength are 
approximately $v_z \propto \left( \Omega_F^2 B_0^2 /\dot{M}_w \right)^{1/3} 
\propto B_0^{1/3} \propto E_{mg}^{1/6}$ and $\dot{M}_w \propto B_0 \propto 
E_{mg}^{1/2}$, where $B_0$ is the initial poloidal magnetic field strength and 
$\Omega_F$ is the angular velocity of the field line (the angular velocity of 
the footpoint on the accretion disk). These are consistent with the results of 
1-dimensional steady solutions although the jet and accretion never reach a 
steady state. It is also obtained that the dependence of the mass accretion 
rate ($\dot{M}_a = dM_a/dt$) on the initial field strength is given by 
$\dot{M}_a \propto B_0^{1.4} \propto E_{mg}^{0.7}$.

In addition to these studies, the acceleration and collimation of
the jet have been studied in the steady (e.g., Sauty, Trussoni, \&
Tsinganos 2004; Bogovalov \& Tsinganos 2005) and nonsteady
framework (e.g., Matsumoto et al. 1996;
Kato et al. 2002; Krasnopolsky, Li, \& Blandford 2003; see also Shibata \&
Kudoh 1999 for a review).
In recent years, interesting simulations taking 
other physical processes, e.g., the magnetic diffusion (Kuwabara et al. 2000; 
Fendt \& \v{C}emelji\'c 2002; Casse \& Keppens 2002, 2004), the 
dynamo process in the accretion disk (von Rekowski et al. 2003), the radiation 
force (Proga 2003) into consideration have been performed.
Koide, Shibata, \& Kudoh (1999) showed that a magnetically driven jet in 
the general relativistic MHD simulation has the characteristics similar to 
those of the nonrelativistic MHD jet (Shibata \& Uchida 1986). 
These works, however, assume the axisymmetry, so that instabilities 
having non-axisymmetric modes are eliminated. It is very interesting 
whether jets are stable against the non-axisymmetric modes or not. Can jets be 
ejected in the same manner with the axisymmetric cases? The stability of 
propagating jet has been studied by taking the 
accretion disk as the fixed boundary condition in 2-D MHD (e.g., Ustyugova et 
al. 1995; Ouyed \& Pudritz 1997a,b, 1999; Romanova et al. 1998; Hardee \& 
Rosen 1999, 2002; Frank et al. 2000), 3-D MHD (Nakamura, Uchida, \& Hirose 
2001; Ouyed, Clarke, \& Pudritz 2003; Nakamura \& Meier 2004), and 3-D SPMHD 
(smoothed particle MHD) simulations (Cerqueira \& de Gouveia Dal Pino 2001). 
The 3-dimensional treatment is necessary for investigating the 
stability of the jet against the non-axisymmetric modes.

Magnetically driven jets have a helical magnetic field because accretion 
disks twist the magnetic field by their own rotation. The generated toroidal 
magnetic field propagates along the magnetic field as torsional Alfv\'en 
waves. It is equivalent that current flows in the jets. The current may cause 
a current-driven instability and jets may be deformed. Nakamura et al. (2001) 
showed that the helical kink instability can explain 
the wiggled structures which are often observed (Krichbaum et al. 1990; 
Hummel et al. 1992; Conway \& Murphy 1993; Roos, Kaastra, \& Hummel 1993; 
Hutchison, Cawthorne, \& Gabuzda 2001; Stirling et al. 2003). The MHD model 
can also explain some characteristics of polarization observations of AGN 
jets. Some AGN jets show 'spine+sheath' $\bm{B}$-field structures, which 
consist of the transverse field in the central region and the longitudinal 
field becoming dominant in some regions offset from the jet axis; it is 
thought that the transverse field is the result of shock compression and that 
the longitudinal field is due to the effect of shear between a jet and a 
surrounding medium (e.g., Attridge, Roberts, \& Wardle 1999). Gabuzda (2003), 
however, pointed out that jets dominated by toroidal field can display the 
'spine+sheath' structure. Uchida et al. (2004) showed this point numerically 
(see Figure 5a and 6a of Uchida et al. (2004)). Besides this, rotation measure 
gradients perpendicular to the jet axis (Asada et al. 2002; Gabuzda, 
Murray, \& Cronin 2004; Zavala \& Taylor 2005) and a distribution of rotation 
measure associated 
with the jet deformation (Feretti et al. 1999) are reproduced numerically by 
the MHD model (Uchida et al. 2004; Kigure et al. 2004). It can be said that 
these points indicate advantages of the MHD model.

Only a few 3-dimensional MHD simulations of jet formation with 
solving the accretion disk self-consistently have been presented (e.g., 
Matsumoto \& Shibata 1997; Matsumoto 1999; Matsumoto \& Shibata 1999; 
Kigure et al. 2002) although 
global (not adopting the shearing box approximation) 3-dimensional MHD 
simulations of the accretion disk around the black hole are vigorously done in 
recent years (e.g., Hawley 2000; Hawley \& Krolik 2001; Machida \& Matsumoto 
2003). In this paper, we study the stability of 
the jet launched from the accretion disk by 3-dimensional MHD simulations. 
We investigate the connection between the non-axisymmetric structure in 
the jet and that in the disk, and compare the characteristics (e.g., the 
dependence of the jet velocity on the magnetic field strength) of jets found 
in 3-dimensional simulations with those of jets found in 2.5-dimensional 
simulations and in steady state models.
In \S 2, we describe the numerical method and the initial and boundary 
conditions. In \S 3, we present results of our simulations. A discussion and 
conclusions are in \S 4 and \S5.

\section{Numerical Method}

\subsection{Assumptions and Basic Equations}
We solve the following ideal MHD equations numerically:
\begin{eqnarray}
\frac{\partial \rho}{\partial t} + \bm{v} \cdot \nabla \rho = -\rho \nabla 
\cdot \bm{v}\\
\frac{\partial \bm{v}}{\partial t} + \bm{v} \cdot \nabla \bm{v} = -\frac{1}
{\rho} \nabla \left( p + \frac{\bm{B}^2}{8 \pi} \right) + \frac{1}{4 \pi \rho} 
\bm{B} \cdot \nabla \bm{B} + \bm{g}\\
\frac{\partial p}{\partial t} + \bm{v} \cdot \nabla p = -\gamma p \nabla 
\cdot \bm{v}\\
\frac{\partial \bm{B}}{\partial t} + \nabla \times \bm{E}= 0\\
\bm{E} = -\bm{v} \times \bm{B}
\end{eqnarray}
where $\rho, p, \bm{v}$ are the density, pressure, and velocity of the gas 
respectively. $\bm{B}$ and $\bm{E}$ are the magnetic and electric field. 
$\gamma$ represents the ratio of specific heats and is equal to 5/3 in this 
paper. The gravity is assumed to be only due to a point-mass gravitational 
potential. This means that the gravitational acceleration, $\bm{g}$, is equal 
to $-\nabla \Psi$, where $\Psi = -GM/(r^2 + z^2)^{1/2}$. $G$ is the 
gravitational constant and $M$ is the mass of a central object.

\subsection{Initial Conditions}
As an initial condition, we assume that an equilibrium disk rotates in 
a central point-mass gravitational potential (e.g., Matsumoto et al. 1996, 
Kudoh et al. 1998). The distributions of angular momentum and pressure are 
assumed as follows (e.g., Abramowicz, Jaroszynski, \& Sikora 1978): 
\begin{eqnarray}
L = L_0 r^a\\
p= K \rho^\Gamma = K \rho^{1+1/n}
\end{eqnarray}
Under these simplifying assumptions, exact solution for the distribution of 
disk material is given by 
\begin{equation}
\psi = -\frac{GM}{(r^2 + z^2)^{1/2}} + \frac{1}{2(1 - a)}L_0^2 r^{2a - 2} + 
(n + 1)\frac{p}{\rho} = const\label{EQdisk}
\end{equation}
From the value of $\psi$ at $(r, z)=(r_0, 0)$, the density and pressure 
distributions in the disk are derived. 
In this paper, $a$ and $n$ are fixed at 0 and 3 respectively.

It is also assumed that there exists a corona outside the disk with uniformly 
high temperature. The corona is in hydrostatic equilibrium without rotation.
The density distribution of the corona is 
\begin{equation}
\rho = \rho_c \rm{exp} \left[ \alpha \left\{ \frac{r_0}{(r^2 + z^2)^{1/2}} - 1 
\right\} \right]
\end{equation}
where $r_0$ is the unit length and equal to $\left( L_0^2/GM \right)
^{1/(1 - 2a)}$. At the point where $(r, z) = (r_0, 0)$, the density of the 
disk is maximum. The parameter $\alpha$ is defined as $\left( \gamma V_{K0}^2 
/V_{sc}^2 \right)$, where $V_{sc}$ is the sound velocity in the corona, 
$V_{K0} = (GM/r_0)^{1/2}$ is the Keplerian velocity at radius $r_0$. 
$\rho_c$ is coronal density at radius $r_0$. $\alpha$ and $\rho_c/\rho_0$ are 
equal to $1.0$ and $10^{-3}$ throughout this paper, where $\rho_0$ is the 
initial density at $(r, z) = (r_0, 0)$. 
To distinguish the material that are initially in the disk from that in the 
corona, a scalar variable, $\Theta$, is introduced. The initial distribution 
of $\Theta$ is as follows: 
\begin{eqnarray}
\Theta = \left\{ \begin{array}{ll}
1& \rm{inside\ of\ the\ disk}\\
0& \rm{outside\ of\ the\ disk}\\
\end{array}\right.
\end{eqnarray}
The inside of the disk is defined as the region where the density obtained by 
equation (\ref{EQdisk}) is positive. The time evolution of $\Theta$ is 
followed by the equation,
\begin{equation}
\frac{d \Theta}{dt} = \frac{\partial \Theta}{\partial t} + v_r 
\frac{\partial \Theta}{\partial r} + \frac{v_\phi}{r} \frac{\partial \Theta}
{\partial \phi} + v_z \frac{\partial \Theta}{\partial z} = 0
\end{equation}
by using the velocity obtained by solving the MHD equations.

The initial magnetic field is assumed to be 
uniform and parallel to the rotation axis of the disk; $\left( B_r, B_\phi, 
B_z \right) = \left( 0, 0, B_0 \right)$.

\subsection{Non-axisymmetric Perturbations in the Disk}
To investigate the stability of the disk and jet system, we add the 
non-axisymmetric perturbation. Two types of perturbations are adopted: 
Either sinusoidal or random perturbation is imposed on the rotational velocity 
of the accretion disk. In the sinusoidal perturbation case, $\delta v_\phi = 
0.1 V_{s0} \sin 2 \phi$ ($0.1 V_{s0}$ is about 3\% of the Keplerian velocity), 
where $V_{s0}$ is the sound velocity 
at $(r,z) = (r_0,0)$ (see Matsumoto \& Shibata 1997, Kato 2002). In the random 
perturbation case, the sinusoidal function in the above-mentioned $\delta 
v_\phi$ is replaced with random numbers between -1 and 1. The case in which 
no perturbation is imposed is also calculated for a comparison.

\subsection{Boundary Conditions}
We impose symmetry for $\rho, v_r, v_\phi, B_z$, and $p$ but anti-symmetry for 
$v_z, B_r$, and $B_\phi$ on the equatorial plane ($z = 0$). The side 
($r=r_{max}$) and top ($z = z_{max}$) surfaces are free boundaries for each 
quantity $Q$ as follows (Shibata 1983): 
\begin{eqnarray}
\frac{\partial \Delta Q}{\partial r} = \frac{\partial \Delta Q}{\partial z} 
= 0\\
\Delta Q = Q(r, \phi, z, t + \Delta t) - Q(r, \phi, z, t)
\end{eqnarray}
On the central axis ($r = 0$), the values of all physical quantities are 
calculated as the average of the points where $r = \Delta r$ ($\Delta r$ is 
the grid spacing in the $r$-direction). The $r$- and $\phi$-components of 
velocity and magnetic field are converted to the $x$- and $y$-components, and 
then the values are averaged. Therefore the velocity and the magnetic field 
can have non-zero $x$- and $y$-components on the axis. In order to avoid a 
singularity at the origin, the region around the origin is treated by 
softening the gravitational potential as 
\begin{eqnarray}
\Psi = \left\{ \begin{array}{ll}
-\frac{GM}{(r^2 + z^2)^{1/2}}&\rm{for}\ \epsilon < (r^2 + z^2)^{1/2}\\
-GM[\frac{1}{\epsilon} - \frac{(r^2 + z^2)^{1/2} - \epsilon}{\epsilon^2} ]
&\rm{for}\ 0.5\epsilon < (r^2 + z^2)^{1/2} \leq \epsilon\\
-\frac{1.5GM}{\epsilon}&\rm{for}\ (r^2 + z^2)^{1/2} \leq 0.5\epsilon\\
\end{array}\right.
\end{eqnarray}
where $\epsilon$ is equal to $0.2 r_0$ throughout this paper.

\subsection{Numerical Method}
The numerical simulations have been carried out using the CIP (Constrained 
Interpolation Profile)-MOC-CT (Method of Characteristics-Constrained 
Transport) scheme. 
The magnetic induction equation is solved by MOC-CT (Evans \& Hawley 1988; 
Stone \& Norman 1992) and the others are solved by CIP method (Yabe \& Aoki 
1991; Yabe et al. 1991). The summary about 
the CIP-MOC-CT scheme is given in Kudoh, Matsumoto, \& Shibata (1999). This 
scheme has been used in MHD simulations of astrophysical jets (Kudoh \& 
Shibata 1997b; Kudoh et al. 1998; Kudoh, Matsumoto, \& Shibata 
2002a,b; Kato et al. 2002). We develop this scheme to 3-dimensional 
cylindrical code.

All physical quantities are normalized by their typical values those are 
initial value at $(r,z) = (r_0,0)$. The  normalized unit for each variable 
is summarized in Table \ref{TABnorm}. For settling the initial condition, 
there are two nondimensional parameters: 
\begin{eqnarray}
E_{th} = \frac{V_{s0}^2}{\gamma V_{K0}^2}\\
E_{mg} = \frac{V_{A0}^2}{V_{K0}^2}
\end{eqnarray}
where $V_{s0} = (\gamma p_0/\rho_0)^{1/2}$, $V_{A0} = B_0/(4 \pi \rho_0)
^{1/2}$, and $p_0$ is the initial pressure at $(r, z) = (r_0, 0)$. $E_{th}$ 
is fixed at 0.05 throughout this paper. When parameter set $a = 0$ and 
$E_{th} = 0.05$, the disk becomes geometrically thick like a torus.

We use eight values of the parameter of initial magnetic field strength, 
$E_{mg}$, ranging from $1.0 \times 10^{-5}$ to $2.0 \times 10^{-3}$. 
These eight runs are done for each perturbation case (sinusoidal, random, or 
no perturbation case). That is, 24 cases are calculated. Table \ref{TABMandP} 
summarizes the names of models and parameters. We select the runs of $E_{mg} 
= 5.0 \times 10^{-4}$ as typical models and display mainly the results of 
these cases.

In order to investigate the difference of the acceleration and collimation 
between the axisymmetric case and the non-axisymmetric cases, we calculate the 
trajectories of fluid elements that are initially on the same magnetic field 
line, treating them as test particles.

The number of grid points in the simulations reported here is 
($N_{r} \times N_{\phi} \times N_{z}$) $=$ $(173 \times 32 \times 197)$. The 
grid points are distributed non-uniformly in the $r$- and $z$-directions. 
The grid spacing is uniform, ($\Delta r, \Delta z$) $=$ ($0.01, 0.01$) for 
$r \leq 1.0$ and $z \leq 1.0$, and then stretched by 5 \% per each grid step. 
The size of the computational domain is $(r_{max}, z_{max}) = (7.5, 16.7)$.

\section{Numerical Results}

\subsection{Time Evolution}
\begin{figure}
\figurenum{1}
\epsscale{0.9}
\plotone{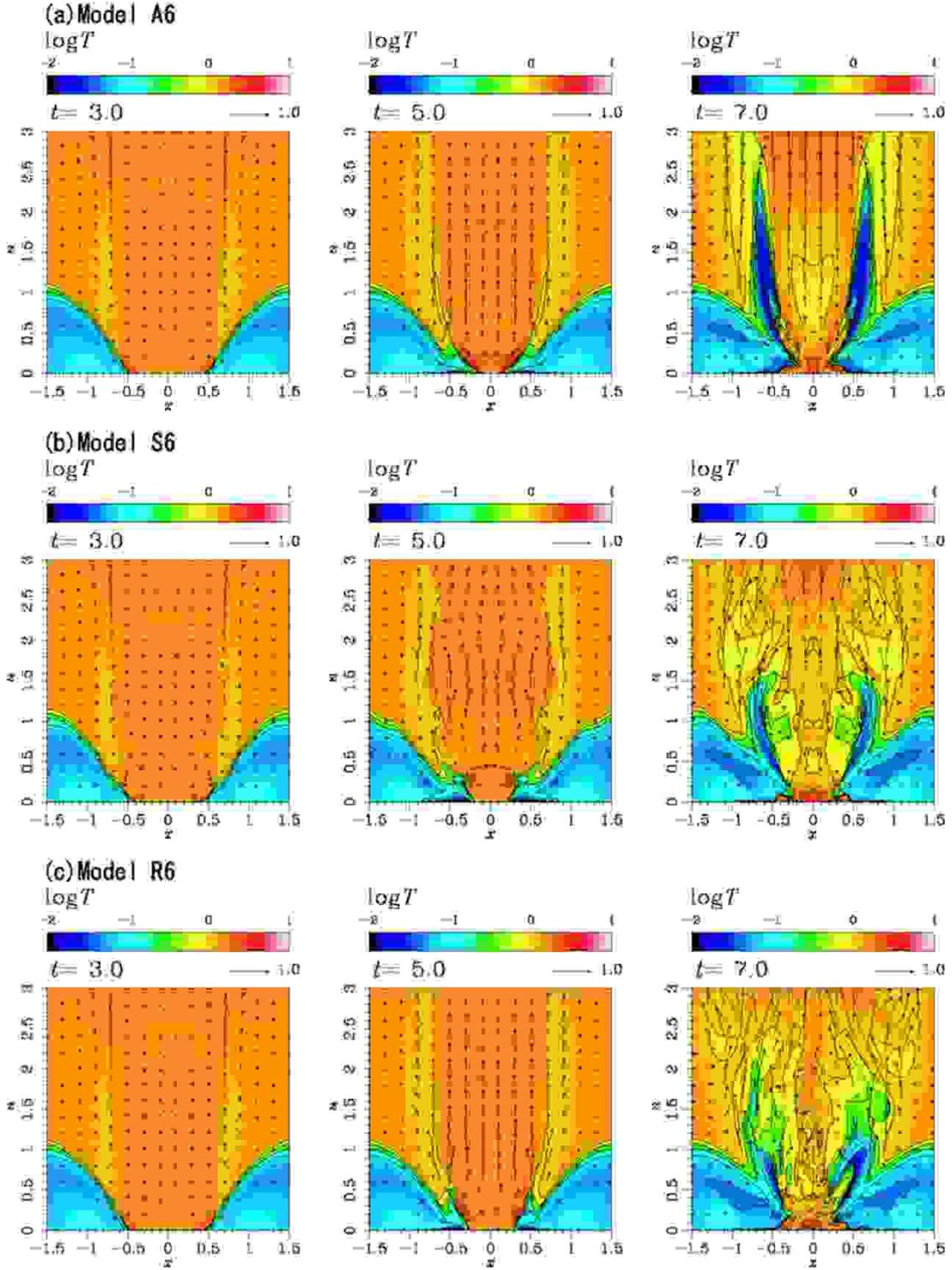}
\caption{Time evolution of the temperature for a model with $E_{mg} = 5.0 
\times 10^{-4}$ in each perturbation case. Time $t \sim 2\pi \simeq 6.28$ 
corresponds to one orbital period at $(r,z)=(1,0)$. Lines show the 
contour of the temperature and arrows show the poloidal velocity.}
\label{FIG01}
\end{figure}

\begin{figure}
\figurenum{2}
\epsscale{0.9}
\plotone{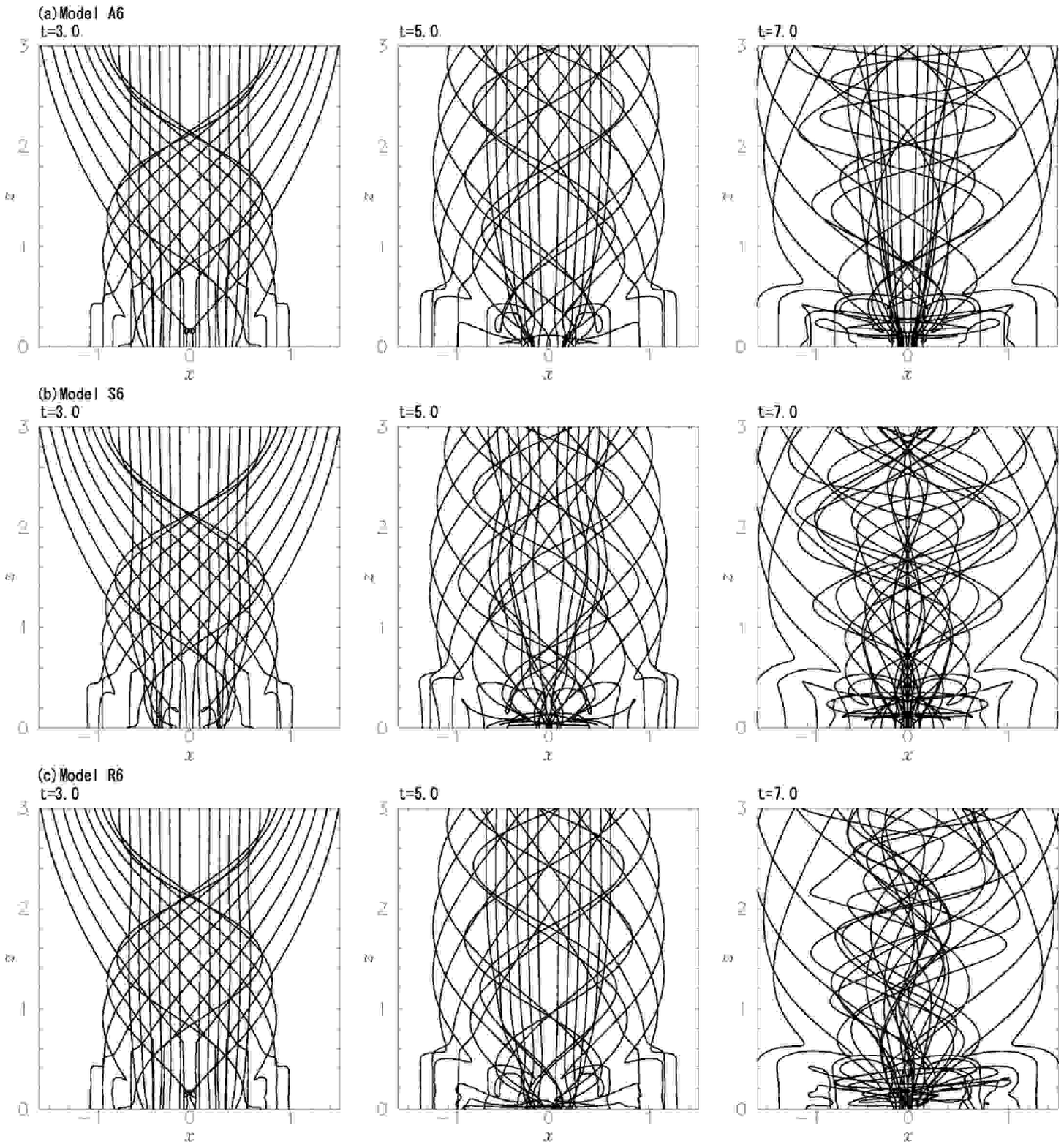}
\caption{Time evolution of the magnetic field lines projected onto the 
$x-z$ plane in each perturbation case.}
\label{FIG02}
\end{figure}

\begin{figure}
\figurenum{3}
\epsscale{0.9}
\plotone{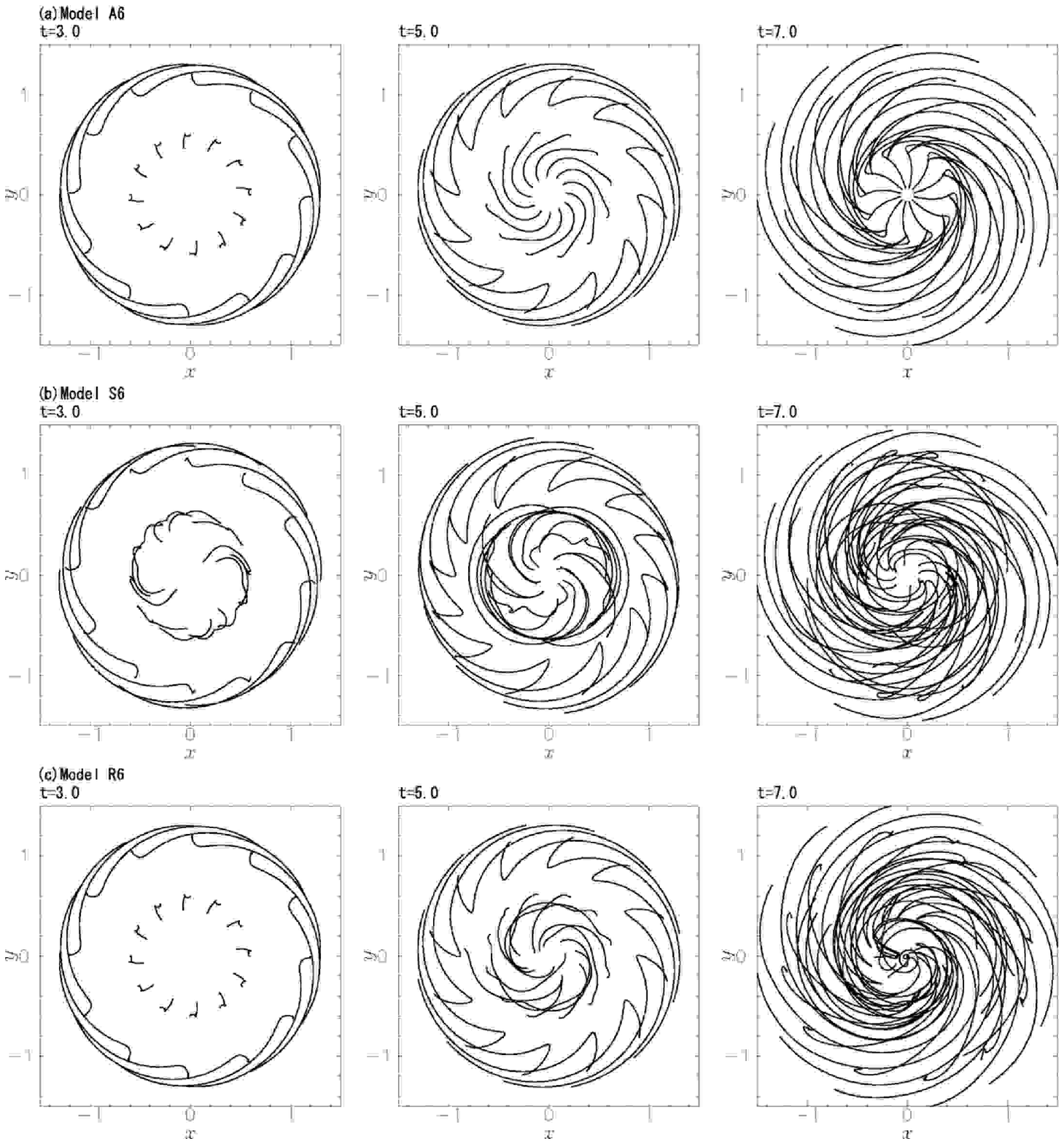}
\caption{Time evolution of the magnetic field lines projected onto the 
$r-z$ plane in each perturbation case.}
\label{FIG03}
\end{figure}

Figure \ref{FIG01} shows the time evolution of the temperature ($T \equiv 
\gamma p / \rho$; the color and contour) and the poloidal velocity (the 
arrows) for the typical model ($E_{mg} = 5.0 \times 10^{-4}$) of each 
perturbation case.
In these cases, the initial values of plasma-$\beta$ ($\equiv 8\pi
p/\bm{B}^2$) are $\beta \simeq 200$ at $(r,z)=(1,0)$, i.e., at the densest 
point of the disk, and $\beta \simeq 4$ at $(r,z)=(0,1)$ in the corona. 
The initial values of plasma-$\beta$ for all models are 
summarized in Table \ref{TABMandP}. Time $t = 2\pi \simeq 6.28$ 
corresponds to one Keplerian orbit at $(r,z)=(1,0)$.
Figure \ref{FIG02} and \ref{FIG03} show the time evolution of selected 
magnetic field lines projected onto the $x-z$ and $x-y$ plane.

The magnetic field is twisted by the disk rotation and the toroidal field is 
continuously generated. It propagates into two directions along the large 
scale magnetic field as a torsional Alfv\'en wave. Deformed magnetic field 
brakes the disk rotation. The disk matter loses the angular momentum and falls 
to the central object while the torsional Alfv\'en wave transports the angular 
momentum to the corona. The rotational velocity of the accreted matter 
increases because the matter falls into the deeper part of the gravitational 
potential according to the 
displacement toward the center. It promotes the 
angular momentum transport by the torsional Alfv\'en wave 
propagation. In the sinusoidal 
perturbation case, the magnetic field is more twisted than in other two cases 
in the early stage ($t=3.0$; see Figure \ref{FIG03}). It indicates more 
effective extraction of the angular momentum from the accretion disk. The 
matter on the disk surface layer falls faster than that in the equatorial 
region in the same way as in Matsumoto et al. (1996) and Kudoh et al. (1998). 
This is called the avalanche flow.

\begin{figure}
\figurenum{4}
\epsscale{1.0}
\plotone{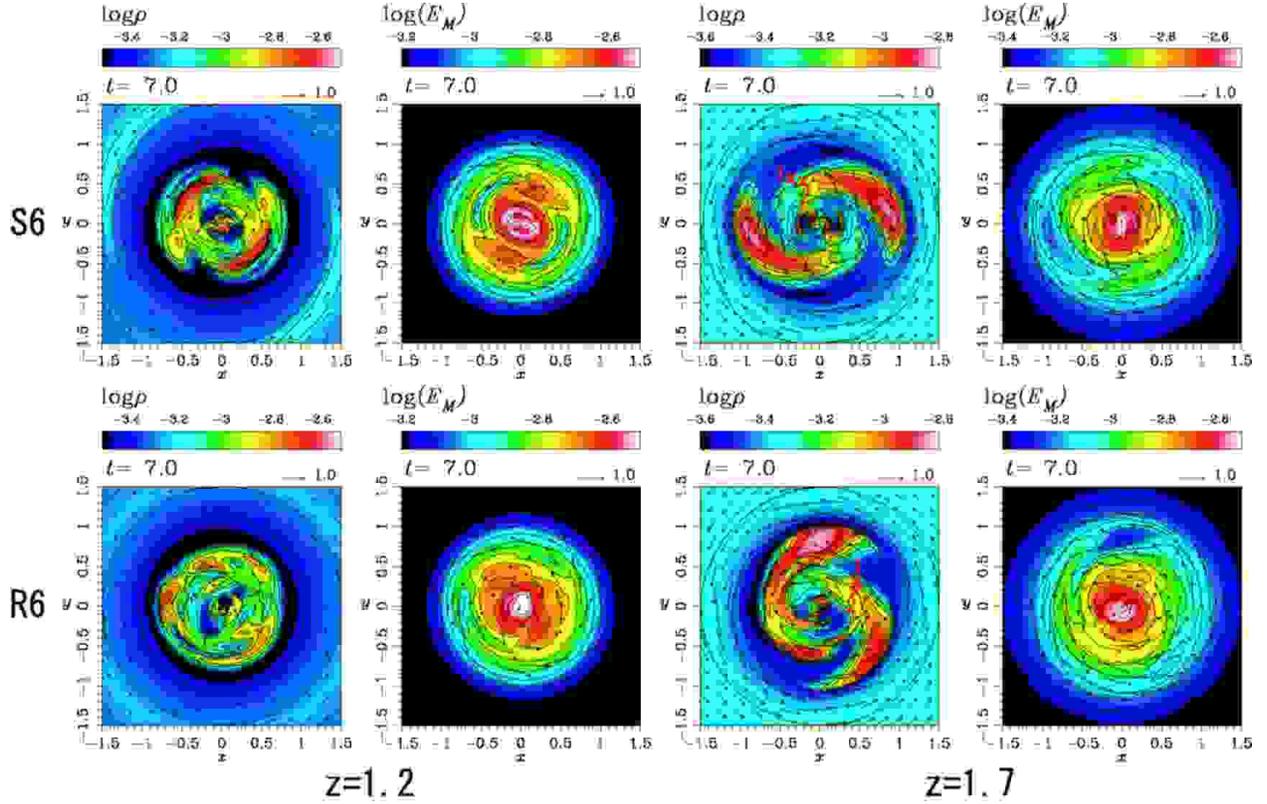}
\caption{The slice images of the jets on the $z=1.2$ and $1.7$ plane. The 
upper four figures are the results of the model S6. The lower four figures are 
the results of the model R6. The color shows the distribution of the density 
or the magnetic energy and arrows show the velocity field projected onto the 
plane. There is an anti-correlation between the density and magnetic field 
distributions. Stability conditions for Kelvin-Helmholtz instability are 
checked between the point $1$ and $2$ in \S4.1.}
\label{FIG04}
\end{figure}

\begin{figure}
\figurenum{5}
\epsscale{0.4}
\plotone{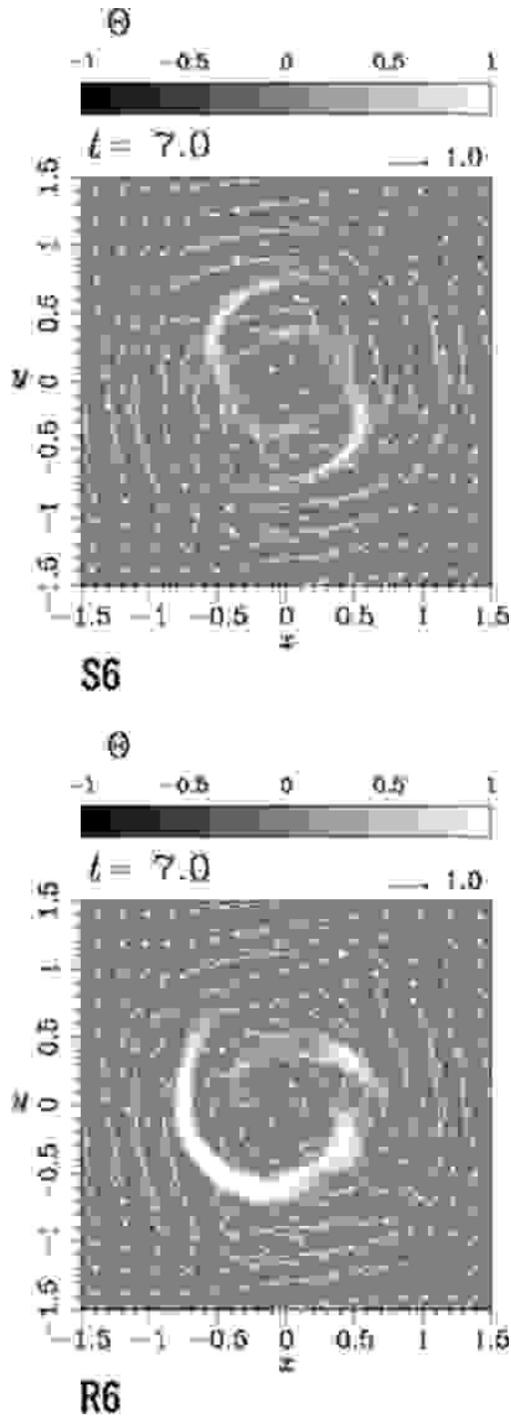}
\caption{The distribution of $\Theta$ on the $z=1.7$ plane at $t=7.0$. 
The region where $\Theta = 1$ means the place the disk matter exists. }
\label{FIG05}
\end{figure}

\begin{figure}
\figurenum{6}
\epsscale{0.7}
\plotone{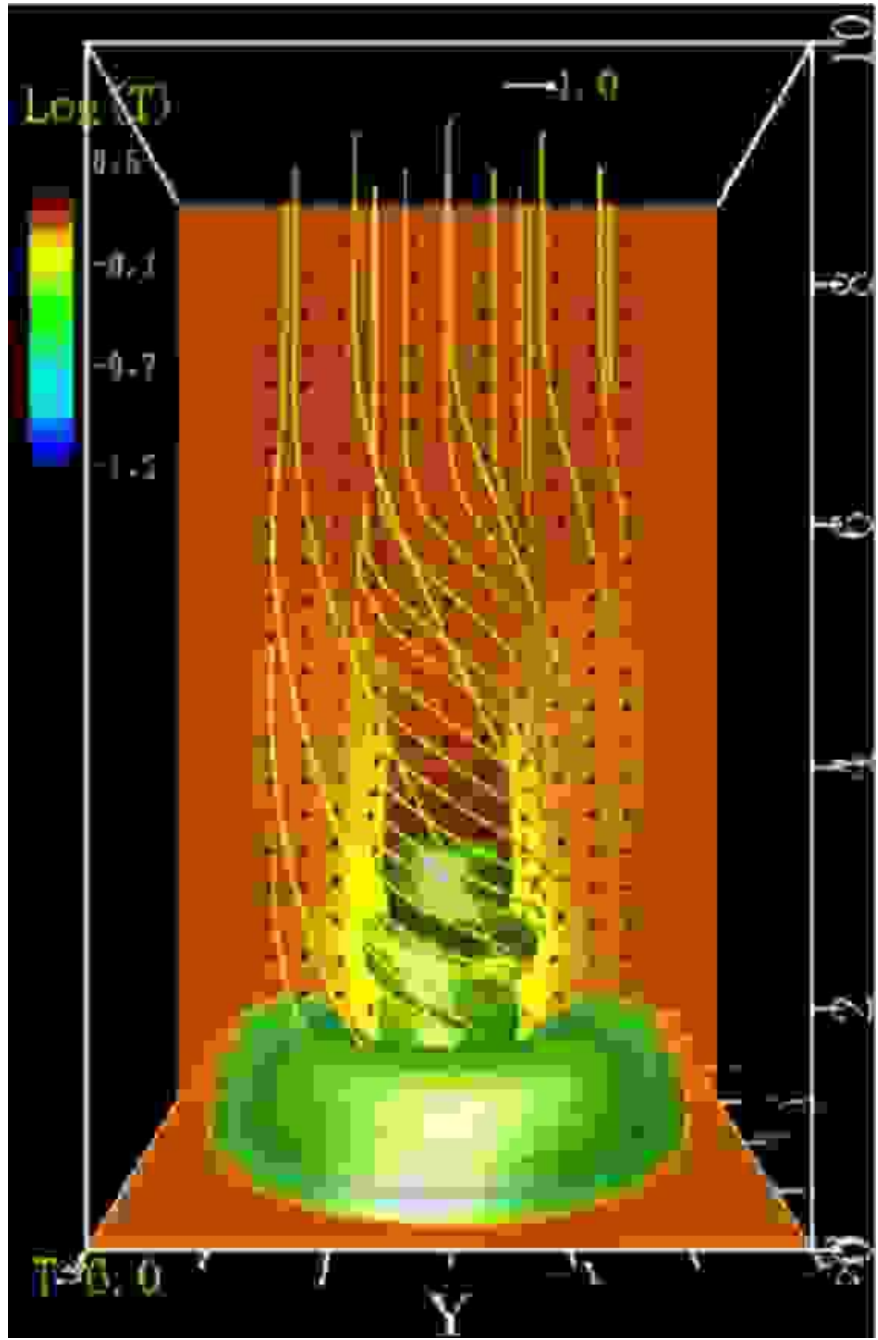}
\caption{The 3-dimensional visualization of the selected magnetic field lines 
and the iso-density surface ($\rho=0.0007$) for the model R6 at $t=6.0$. The 
color shows the distribution of logarithmic temperature and the arrows show 
the velocity field on the $x=0.0$ plane.}
\label{FIG06}
\end{figure}

The disk rotation amplifies the magnetic field and 
the magnetic pressure increases near the equatorial plane. 
The magnetic pressure gradient force as well as the magnetic centrifugal force 
accelerates the disk matter in the 
vertical direction and the jet is formed (see Figure 7d).
Just after the ejection of the jet ($t=5.0$), a disk 
deformation is most remarkable in the sinusoidal perturbation case. 
It is also apparent that stream lines of outflow show bending pattern 
in the sinusoidal case although the difference between the axisymmetric case 
and the random perturbation case is not so notable. 
At $t=7.0$, the difference between the model A6 and the model R6 also becomes 
clear. 
Figure \ref{FIG04} shows the slice images of the jets on the $z=1.2$ and $1.7$ 
planes at $t=7.0$. The upper four figures are the results of the model S6. The 
lower four figures are the results of the model R6. The color shows the 
distribution of the density or the magnetic energy and arrows show the 
velocity field projected onto the plane. There is an anti-correlation between 
the density and magnetic field distributions. Stability conditions for 
Kelvin-Helmholtz instability are checked between the point $1$ and $2$ in 
\S4.1.
In the model S6, initial perturbation is of $m=2$ ($m$ is the 
azimuthal wave number), so that physical variable distributions have the 
periodicity of $\Delta \phi = \pm \pi$. 
It is to be noted that the high density regions in Figure \ref{FIG04} are not 
jets. The high density regions are formed by the compression of the corona 
matter. This is certified by Figure \ref{FIG05} showing the distribution of 
$\Theta$ on the $z=1.7$ plane at $t=7.0$. The region where $\Theta = 1$ means 
the place the disk matter exists. 
Figure \ref{FIG06} is the 3-dimensional visualization of the selected 
magnetic field lines and the iso-density surface for the model R6 at 
$t=6.0$. A helical pattern of the density distribution is seen.


\subsection{The Lagrangian Fluid Elements along a Field Line}
\begin{figure}
\figurenum{7}
\epsscale{0.5}
\plotone{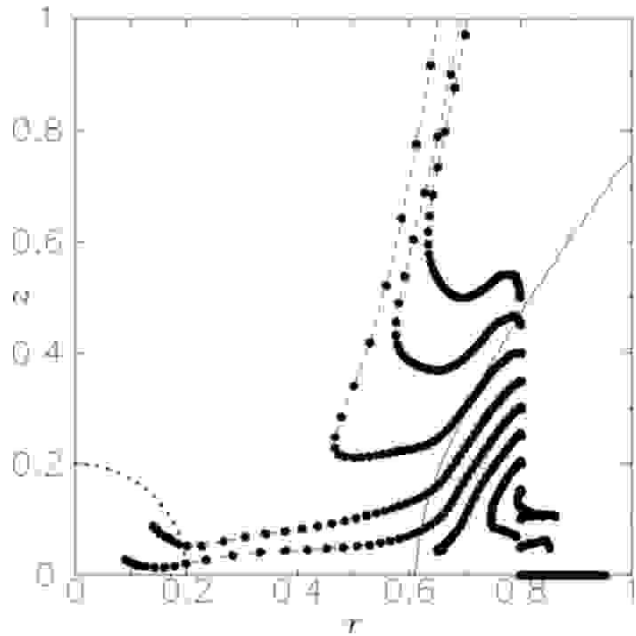}
\caption{The trajectories of the Lagrangian fluid elements $(0 < t < 6.1)$ 
which are initially on a magnetic field line $(r=0.8$ at $t=0)$ in the model 
A6. The solid line shows the initial disk surface. The dotted line indicates 
the region where the gravitational potential is softened $(R = (r^2 + z^2)^
{1/2} < 0.2)$. The dashed line shows the trajectory of each element and the 
filled circles show the Lagrangian fluid elements plotted with time interval 
of 0.1.}
\label{FIG07}
\end{figure}

Figure \ref{FIG07} shows the trajectories of the Lagrangian fluid 
elements $(0 < t < 6.1)$ which are initially on a magnetic field line $(r=0.8$ 
at $t=0)$ in the model A6. 
The elements initially located near the disk 
surface lose the angular momentum and are accreted toward the central object 
or ejected as a jet. Those initially located near the equatorial plane 
remain in the disk.

\begin{figure}
\figurenum{8}
\epsscale{1.0}
\plotone{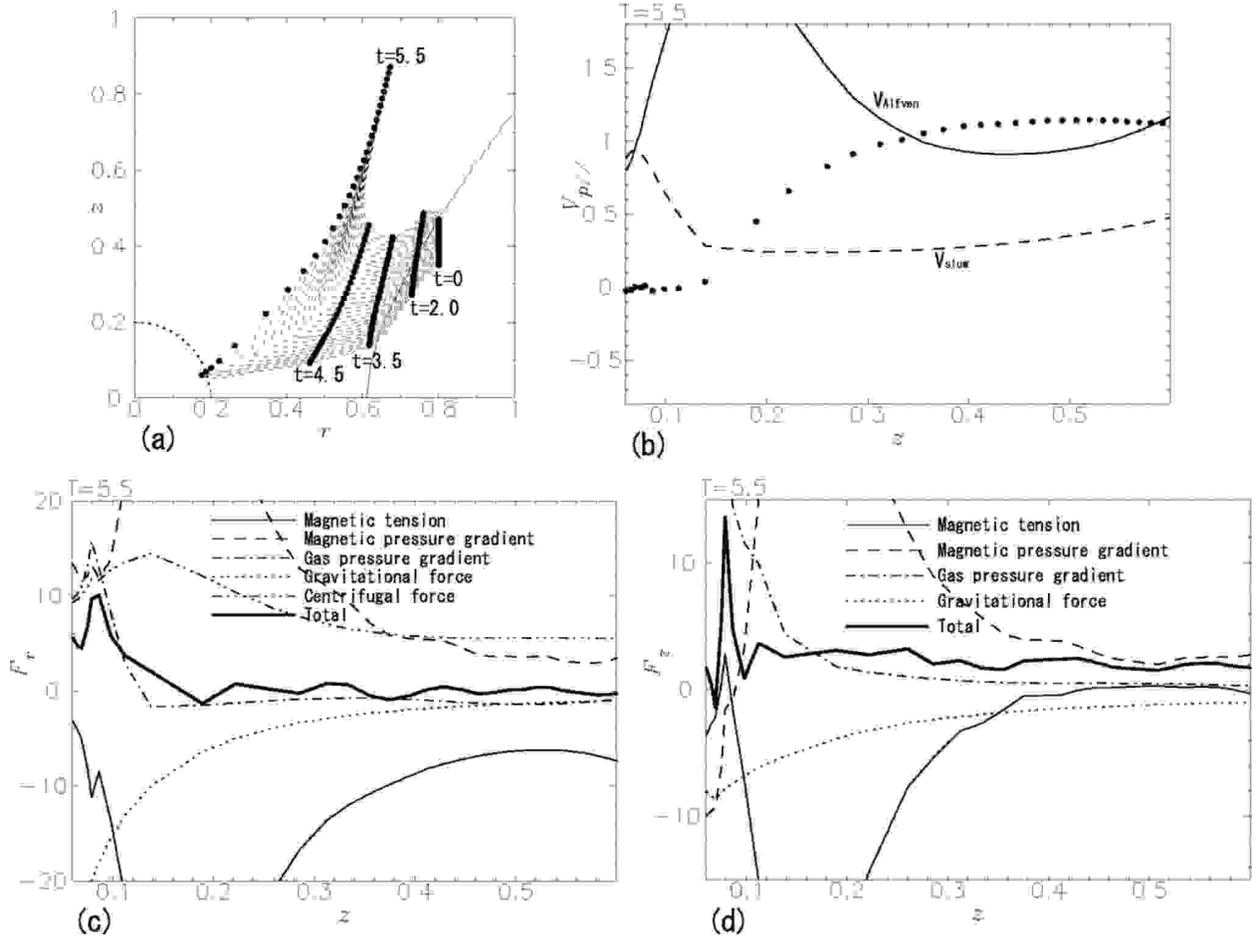}
\caption{(a)The trajectories of the Lagrangian fluid 
elements $(0 < t < 5.5)$ those are initially located near the disk surface 
and on the same magnetic field line as shown in Figure \ref{FIG07}. 
(b)The poloidal velocity along the magnetic field line $(v_{p\parallel})$ at 
the each Lagrangian fluid element position 
at $t=5.5$. 
The horizontal axis is the position ($z$) of each element 
and the vertical axis is the velocity. The solid line shows the poloidal 
Alfv\'en velocity and the dashed line shows the slow magnetosonic velocity. 
(c)The $r$-component of each force. (d)The $z$-component of each force. 
The solid line is the magnetic tension, the 
dashed line is the magnetic pressure gradient force, the dash-dotted line is 
the gas pressure gradient force, the dotted line is the gravitational force, 
the dash-triple-dotted line is the centrifugal force, 
and the thick solid line is the sum of them.}
\label{FIG08}
\end{figure}

Figure \ref{FIG08}a shows the trajectories of the Lagrangian fluid 
elements $(0 < t < 5.5)$ which are initially located near the disk surface 
and on the same magnetic field line as shown in Figure \ref{FIG07}. 
Figure \ref{FIG08}b indicates the poloidal velocity along the magnetic 
field line $(v_{p\parallel})$ at each Lagrangian fluid element position 
at $t=5.5$. 
The horizontal axis is the position ($z$) of each element 
and the vertical axis is the velocity. The solid line shows the poloidal 
Alfv\'en velocity and the dashed line shows the slow magnetosonic velocity. 
The elements are accelerated to be superslow magnetosonic and 
trans-Alfv\'enic. Figure \ref{FIG08}c and \ref{FIG08}d display the $r$- and 
$z$-components of each force. The solid line is the magnetic tension, the 
dashed line is the magnetic pressure gradient force, the dash-dotted line is 
the gas pressure gradient force, the dotted line is the gravitational force, 
the dash-triple-dotted line is the centrifugal force, 
and the thick solid line is the sum of them. 
The magnetic tension (hoop stress) balances with the magnetic pressure 
gradient force + centrifugal force, and maintains the collimated flow. 
The magnetic pressure gradient force and the centrifugal force accelerate the 
flow in the axial ($z$) and radial ($r$) 
directions. The gas pressure gradient force tends to zero and becomes 
inefficient.

\begin{figure}
\figurenum{9}
\epsscale{0.5}
\plotone{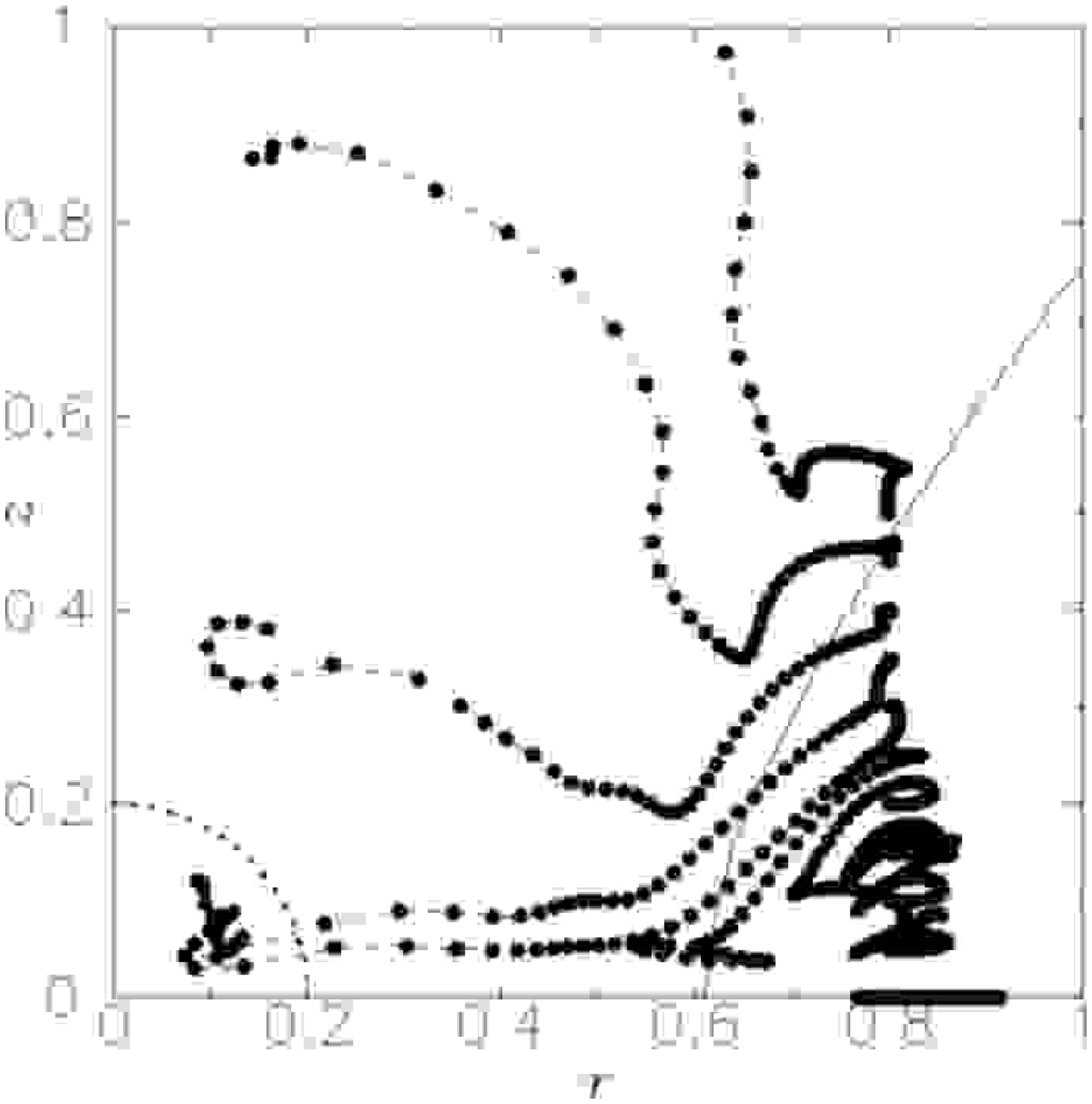}
\caption{Similar to Fig. \ref{FIG07}, but for the model S6. Initial azimuthal 
position of the Lagrangian fluid elements is 0.}
\label{FIG09}
\end{figure}

\begin{figure}
\figurenum{10}
\epsscale{1.0}
\plotone{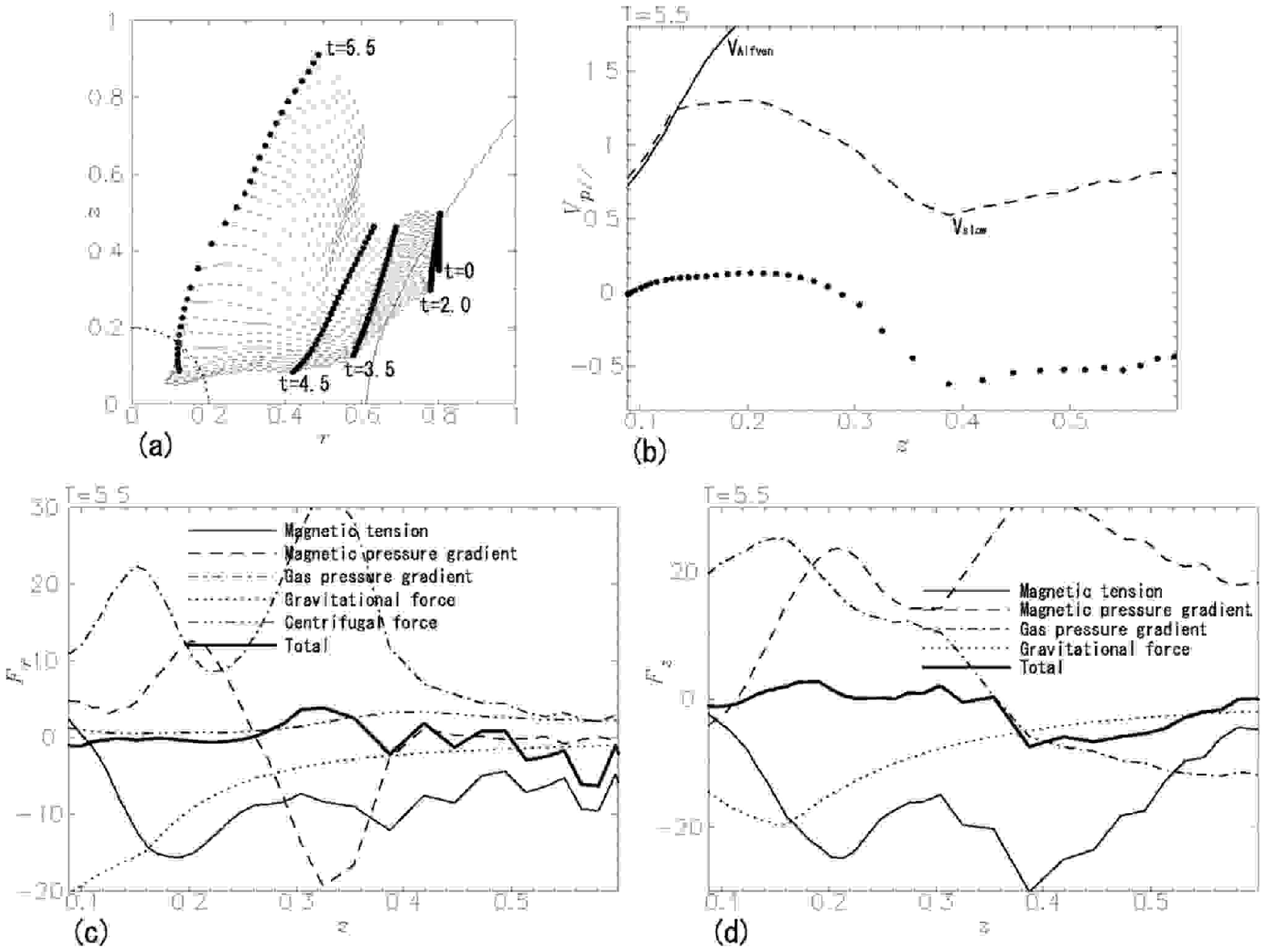}
\caption{Similar to Fig. \ref{FIG08}, but for the model S6. Initial azimuthal 
position of the Lagrangian fluid elements is 0.}
\label{FIG10}
\end{figure}

\begin{figure}
\figurenum{11}
\epsscale{0.5}
\plotone{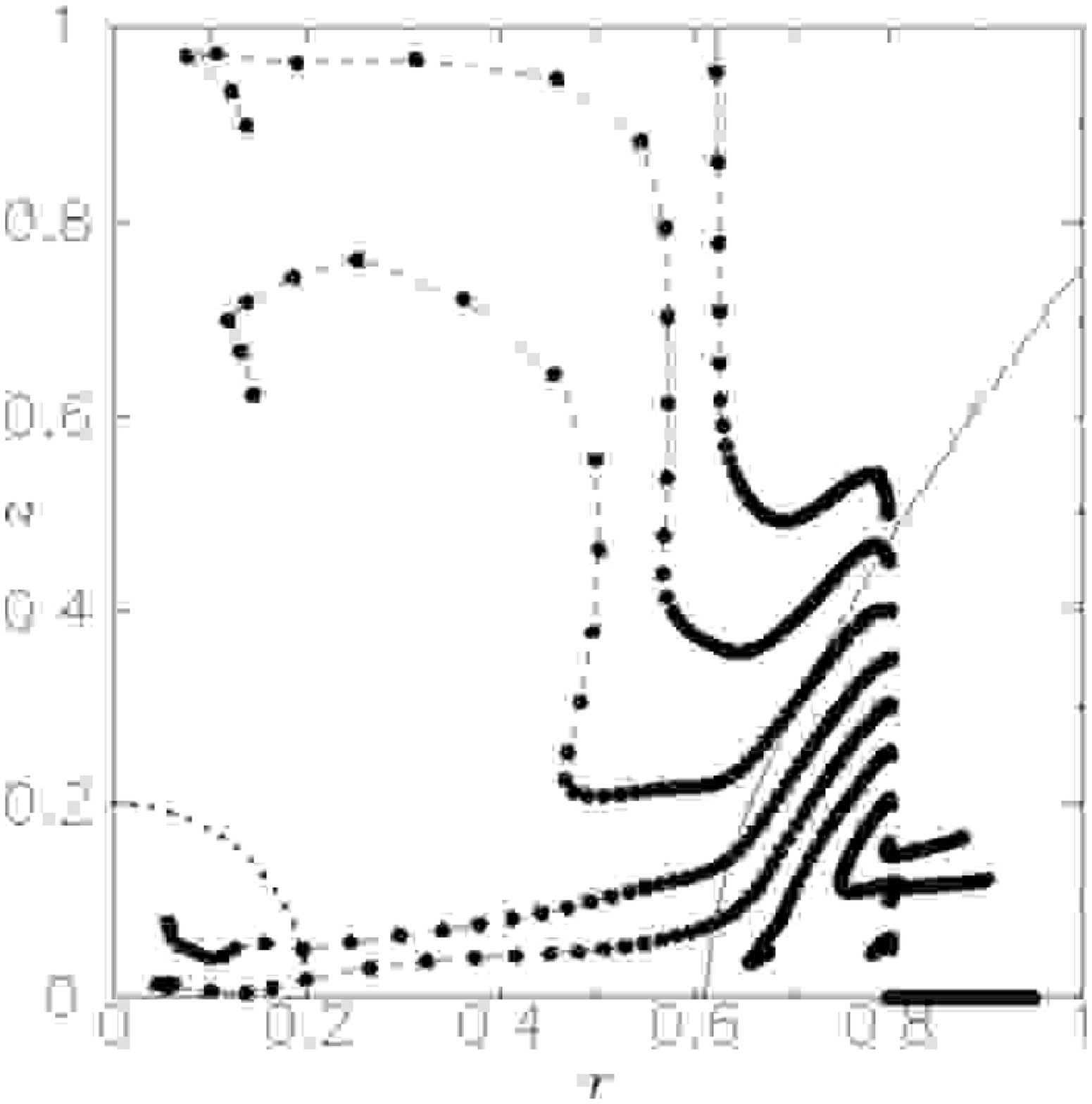}
\caption{Similar to Fig. \ref{FIG07}, but for the model R6. Initial azimuthal 
position of the Lagrangian fluid elements is 0.}
\label{FIG11}
\end{figure}

\begin{figure}
\figurenum{12}
\epsscale{1.0}
\plotone{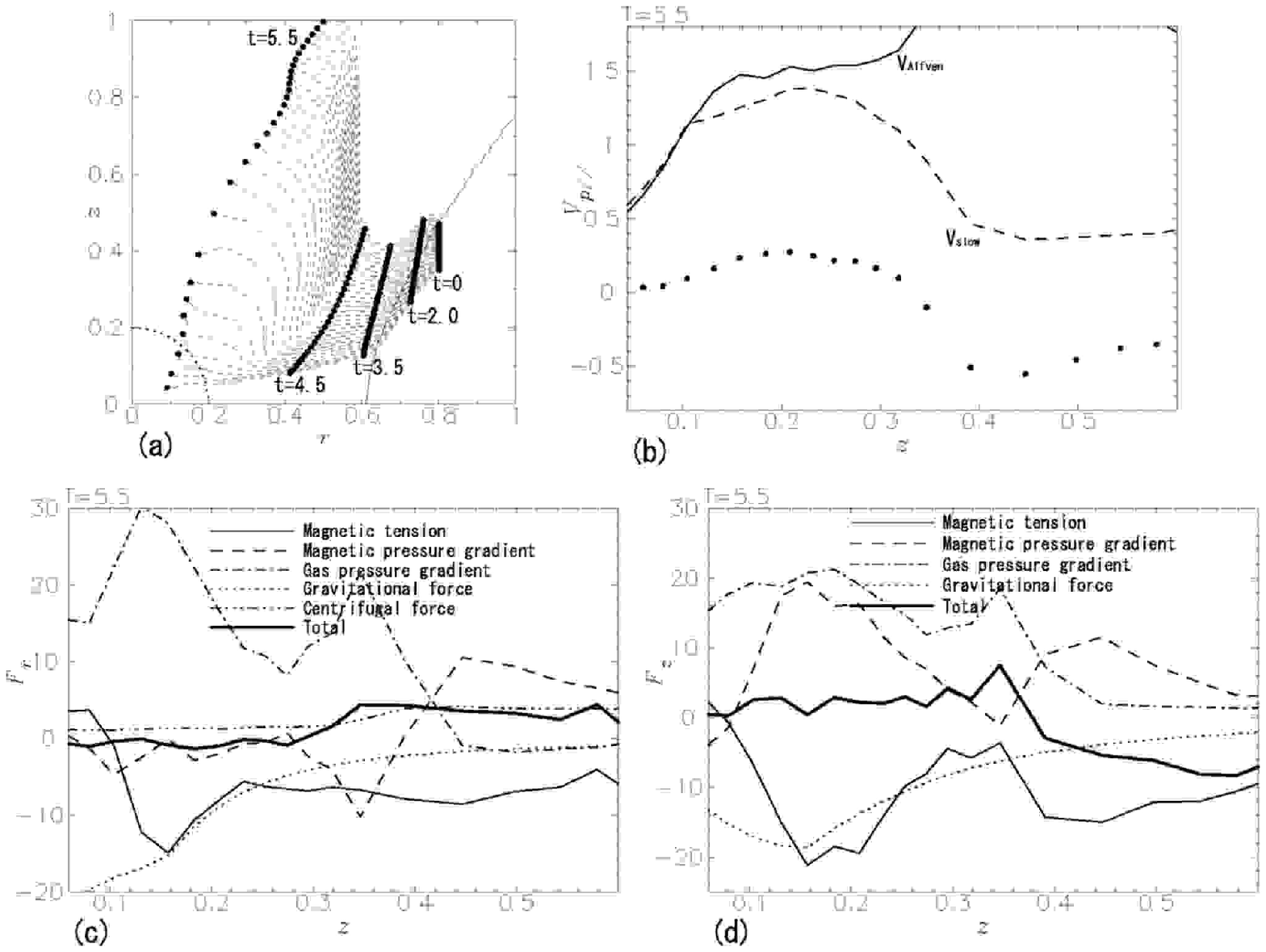}
\caption{Similar to Fig. \ref{FIG08}, but for the model R6. Initial azimuthal 
position of the Lagrangian fluid elements is 0.}
\label{FIG12}
\end{figure}

In the non-axisymmetric calculations, the trajectories are different according 
to the initial azimuthal position of the Lagrangian fluid elements. A part of 
elements show trajectories similar to those in the axisymmetric calculation, 
but another part of elements show remarkably different trajectories. 
Figure \ref{FIG09} is similar to Figure \ref{FIG07}, but for the model S6. 
Initial azimuthal position of the elements is $\phi = 0$. Part of elements 
whose initial $z$ positions are the same as those ejected as a jet in the 
model A6 
move toward the rotational axis of the accretion disk. This is caused because 
the radial magnetic pressure gradient and centrifugal forces against the 
magnetic pinch force are smaller than those in the axisymmetric case (see 
Figure \ref{FIG10}c). 
Figure \ref{FIG10}a shows the trajectories of the elements 
initially located near the disk surface and on the same magnetic field line 
as shown in Figure \ref{FIG09}. A part of the elements are once accelerated in 
the $z$-direction (after $t=4.5$), but those are forced to go toward the axis. 
The angle between the poloidal velocity and the poloidal magnetic field is 
larger than $\pi/2$ (see Figure \ref{FIG10}b). Figure \ref{FIG10}c and 
\ref{FIG10}d indicate the $r$- and $z$-components of each force. The 
$r$-components of the magnetic and gas pressure gradient forces show 
variabilities in the $z$-direction compared to the model A6. The gas pressure 
gradient force does not tend to zero in both $r$- and $z$-components. 
The $r$-component of the magnetic tension shows the bending curve. This may 
be the reason of the bending pattern of the jet (see \S 3.1 and Figure 
\ref{FIG01}). 
Figure \ref{FIG11} and \ref{FIG12} are similar to Figure \ref{FIG07} and 
\ref{FIG08} for the model R6. The reason for a element going toward the 
rotational axis is the same as for the model S6; that is, the decrease of the 
radial magnetic pressure gradient and centrifugal forces. Initial azimuthal 
position of the 
elements is 0. These figures indicate the characteristics similar to those 
for the model S6 except the gas pressure gradient force being nearly equal to 
zero beyond $z \sim 0.45$.

\subsection{Dependences on the Initial Magnetic Field Strength}
\begin{figure}
\figurenum{13}
\epsscale{0.5}
\plotone{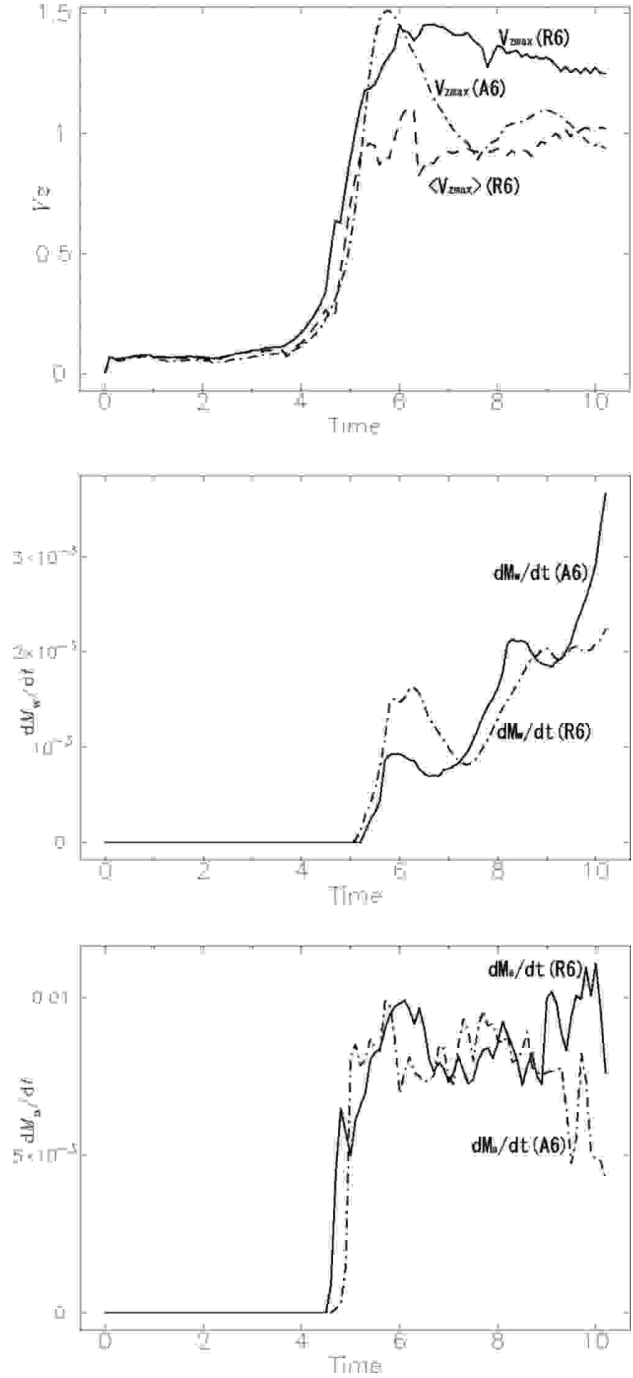}
\caption{The time variation of the vertical velocity ($v_z$), the mass outflow 
rate, and the mass accretion rate for the model R6 and A6. 
The mass outflow rate is measured on the $z=1.0$ plane and the mass accretion 
rate is measured on the $r=0.2$ plane (see \S3.3). 
As for the vertical velocity, the spatially maximum value of $v_z$ and 
azimuthally averaged $v_z$ ($\langle v_z \rangle$) for the model R6 are 
plotted.}
\label{FIG13}
\end{figure}

Kudoh et al. (1998) pointed out that the characteristics of the jet and 
accretion in nonsteady MHD simulations are similar to those of the steady 
ones, although the jet and disk in the simulations are essentially nonsteady. 
(In the resistive cases, Kuwabara et al. (2000) and Casse \& Keppens (2002, 
2004) showed that the jet and 
accretion reach the steady state.) How about in the 3-dimensional cases? 
The jet and accretion in the 3-dimensional cases are also nonsteady. 
As an example, we show the time variation of the vertical velocity ($v_z$), 
the mass outflow rate, and the mass accretion rate 
for the model R6 and A6 (Figure \ref{FIG13}). 
As for the vertical velocity, the spatially maximum value of $v_z$ and 
azimuthally averaged $v_z$ ($\langle v_z \rangle$) for the model R6 are 
plotted.

To investigate the macroscopic characteristics of the non-axisymmetric MHD 
jets, the dependences of the maximum jet velocities $(v_z)$, the maximum mass 
outflow rates $(\dot{M}_w = dM_w/dt)$, and the maximum mass accretion rates 
$(\dot{M}_a = dM_a/dt)$ on the initial magnetic field strength, $E_{mg}$, are 
calculated and compared with those in the axisymmetric case. These values are 
measured only for the material that is initially in the disk by using the 
function $\Theta$.

\begin{figure}
\figurenum{14}
\epsscale{1.0}
\plotone{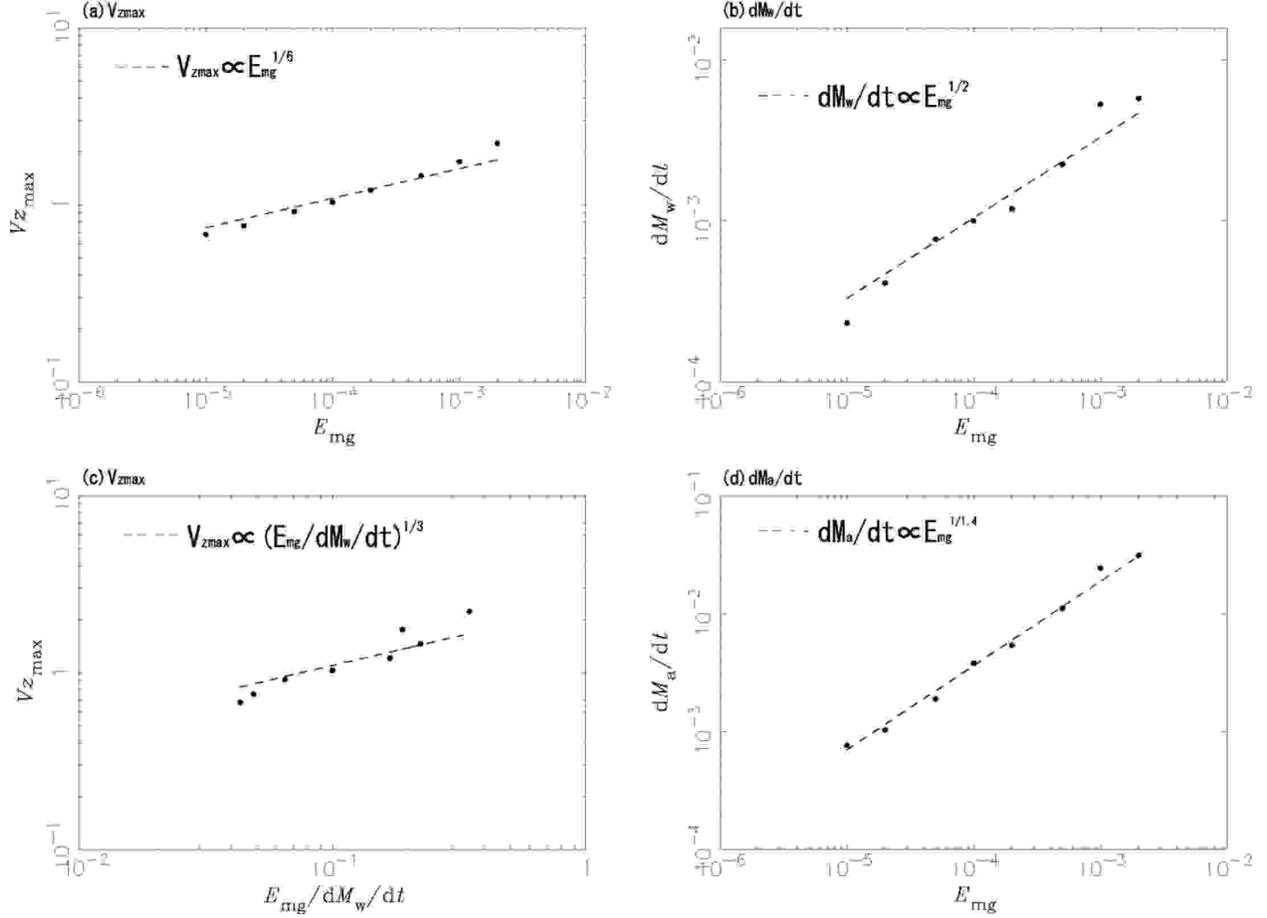}
\caption{The dependences on the initial magnetic field strength for the random 
perturbation cases. (a)The maximum velocities of jets as a function of 
magnetic energy $E_{mg} = V_{A0}^2/V_{K0}^2$ for the random perturbation 
cases. The broken line shows $v_z \propto E_{mg}^{1/6}$. (b)The maximum mass 
outflow rates of jets as a function of magnetic energy $E_{mg} = V_{A0}^2/
V_{K0}^2$ for the random perturbation cases. The broken line shows $dM_w/dt 
\propto E_{mg}^{0.5}$. (c)The maximum velocities of jets as a function of 
$E_{mg}/dM_w/dt$ for the random perturbation cases. The broken line shows $v_z 
\propto \left( E_{mg}/dM_w/dt \right)^{1/3}$. (d)The maximum mass accretion 
rates of jets as a function of magnetic energy $E_{mg} = V_{A0}^2/V_{K0}^2$ 
for the random perturbation cases. The broken line shows $dM_a/dt \propto 
E_{mg}^{1/1.4}$.}
\label{FIG14}
\end{figure}

As shown in Kudoh et al. (1998), the maximum jet velocity is proportional 
to $E_{mg}^{1/6} \propto B_0^{1/3}$ and is of the order of the Keplerian 
velocity. The maximum jet velocities in both sinusoidal and random 
perturbation cases have the same dependence as that in the axisymmetric case 
and are also of the order of the Keplerian velocity. As an example, the 
maximum jet velocities as a function of $E_{mg}$ in the random perturbation 
case are shown in Figure \ref{FIG14}a. The broken line shows $v_z \propto 
E_{mg}^{1/6} \propto B_0^{1/3}$.

The definition of the mass outflow rate $(\dot{M}_w)$ of the jet is 
\begin{equation}
\dot{M}_w = \int_{0}^{2\pi} \int_{0}^{1} \rho v_z r dr d\phi\\
\end{equation}
at $z=1$. 
Kudoh et al. (1998) showed that $\dot{M}_w$ is 
proportional to $E_{mg}^{0.5} \propto B_0$. The maximum mass 
outflow rates in both perturbation cases also have the same dependence on 
$E_{mg}$ as that in the axisymmetric case. Figure \ref{FIG14}b shows the 
maximum mass outflow rates as a function of $E_{mg}$ in the random 
perturbation case. The broken line shows $\dot{M}_w \propto E_{mg}^{0.5} 
\propto B_0$.

As mentioned in Kudoh et al. (1998), these two relations are interpreted as 
the Michel's solution, $v_\infty \propto (B_0^2/\dot{M}_w)^{1/3}$, in the 
magnetic pressure driven jet regime ($\dot{M}_w \propto E_{mg}^{0.5} \propto 
B_0$). Figure \ref{FIG14}c shows the maximum jet velocities as a function of 
$E_{mg}/\dot{M}_w$ in the random perturbation case. The broken line shows 
$v_z \propto (E_{mg}/\dot{M}_w)^{1/3}$. The maximum jet velocities in our 
simulations are approximately proportional to $(E_{mg}/\dot{M}_w)^{1/3}$. 
This result indicates that the MHD jets even in the 3-dimension are consistent 
with the Michel's steady solution, even before the jets reaching the steady 
state.

The mass accretion rate $(\dot{M}_a)$ is defined by 
\begin{equation}
\dot{M}_a = -\int_{0}^{2\pi} \int_{0}^{1} \rho v_r r d\phi dz
\end{equation}
at $r=0.2$. 
Figure \ref{FIG14}d shows the maximum mass accretion rates as a 
function of $E_{mg}$ in the random perturbation case. The broken line shows 
$\dot{M}_a \propto E_{mg}^{1/1.4}$. $\dot{M}_a$ is well 
fitted by $\dot{M}_a \propto E_{mg}^{1/1.4} \sim E_{mg}^{0.7} \propto 
B_0^{1.4}$ as same as in the axisymmetric case shown in Kudoh et al. (1998). 
$\dot{M}_a$ in the sinusoidal perturbation case also has the same dependence 
on $E_{mg}$.

\section{Discussion}

\subsection{Non-axisymmetric Structure in the Jets}
Lobanov \& Zensus (2001) found that the 3C273 jet has a double helical 
structure and that it can be fitted by two surface modes and three body modes 
of Kelvin-Helmholtz (K-H) instability. On the other hand, the jet launched 
from the disk in our simulation has a non-axisymmetric ($m=2$) structure 
in both perturbation cases (see Figure \ref{FIG04}). 

The stability condition for non-axisymmetric K-H surface modes is 
\begin{eqnarray}
\Delta V < V_{As} = \left[ \frac{\rho_1 + \rho_2}{4 \pi \rho_1 \rho_2}
\left( {\bm{B}_1}^2 + {\bm{B}_2}^2 \right)\right]^{1/2}
\end{eqnarray}
where $\Delta V \equiv V_1 - V_2$ and $V_{As}$ is called as a ``surface'' 
Alfv\'en velocity (Hardee \& Rosen 2002).
Figure \ref{FIG04} shows the distribution of logarithmic density on the 
$z=1.7$ plane at $t=7.0$ for the model S6 and R6. We check the above-mentioned 
stability condition between the point $1$ and $2$. In the model S6, $\Delta 
V_{12}=0.38$ (in this case, $\Delta V$ is a 
difference of $v_z$) and $V_{As12}=16$. In the model R6, $\Delta 
V_{12}=0.29$ and $V_{As12}=16$. Therefore, the 
non-axisymmetric structure in the jets is not produced by K-H surface modes.

Non-axisymmetric K-H body modes become unstable if jet velocity is 
super-fast magnetosonic ($V_{FM} < V_j$) or is slightly below the slow 
magnetosonic velocity ($C_s V_A/(C_s^2 + V_A^2)^{1/2}  < V_j < V_{SM}$) 
(Hardee \& Rosen 1999), where $C_s$ is the sound velocity and $V_A$ is the 
Alfv\'en velocity. The jets in the model S6 and R6 satisfy the former 
unstable condition for a little time, but after the short time unstable phase 
the jets become stable 
for that condition. It can be said that body modes of K-H instability can not 
explain the production of the non-axisymmetric structure.

As for the velocity field, the observational results indicating the rotation 
of YSO jets have recently been obtained (e.g., Bacciotti et al. 2002; Coffey 
et al. 2004). In our simulations, the jets have the non-axisymmetric helical 
velocity field (see $v_{zmax}$ and $\langle v_{zmax} \rangle$ in Figure 
\ref{FIG13}). Such velocity field may be observed in the future.

\subsection{Connection of Non-axisymmetric Structure between Disk and Jet}
To investigate a connection of the non-axisymmetric structure in the jet and 
that in the disk, we calculate the Fourier power spectra of the 
non-axisymmetric modes of the magnetic energy in the disk or in the jet.
\begin{equation}
\tilde{E}_M \left(k_r,m,k_z\right) = \frac{1}{V_s}\int\!\!\!\int\!\!\!\int_
{V_s}E_M \left(r,\phi,z\right) e^{i\left( k_r r + m \phi + k_z z\right)} rdr 
d\phi dz
\end{equation}
where $E_M$ is the magnetic energy $(\bm{B}^2/8\pi)$. $V_s$ means the volume 
of the disk or the jet. To concentrate on non-axisymmetric modes, 
$\tilde{E}_M$ is integrated about the radial and axial wave number ($k_r$ and 
$k_z$).
The definition of the disk or the jet is the following: 
\begin{itemize}
\item{The disk is the region where the function $\Theta$ is not equal to 0 
and $z \leq$ 1.5.}
\item{The jet is the region where the function $\Theta$ is not equal to 0 
and $z >$ 1.5.}
\end{itemize}

\begin{figure}
\figurenum{15}
\epsscale{1.0}
\plotone{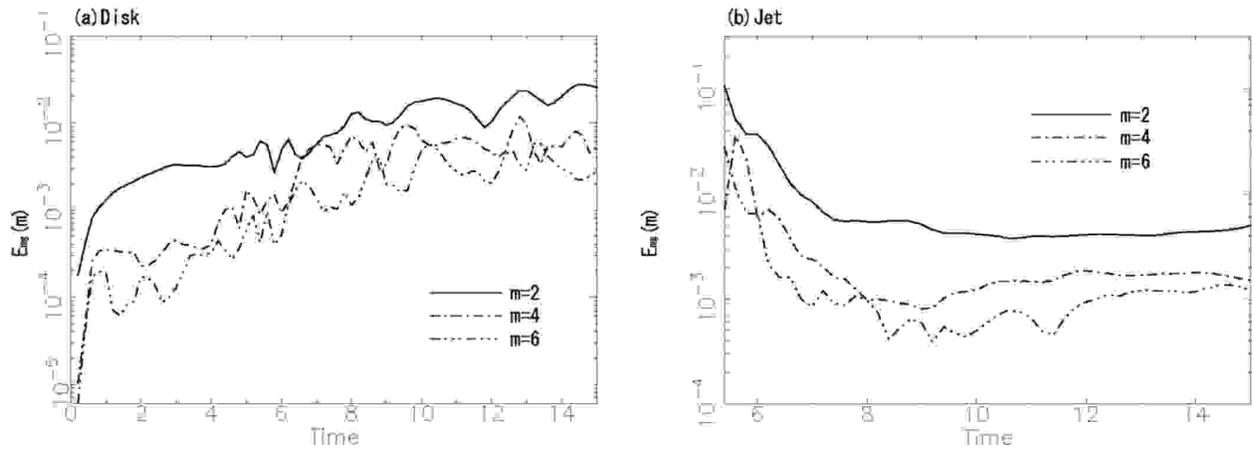}
\caption{Time evolution of Fourier power spectra of the magnetic energy 
in the model S6. (a)In the disk and (b)in the jet.}
\label{FIG15}
\end{figure}

\begin{figure}
\figurenum{16}
\epsscale{0.9}
\plotone{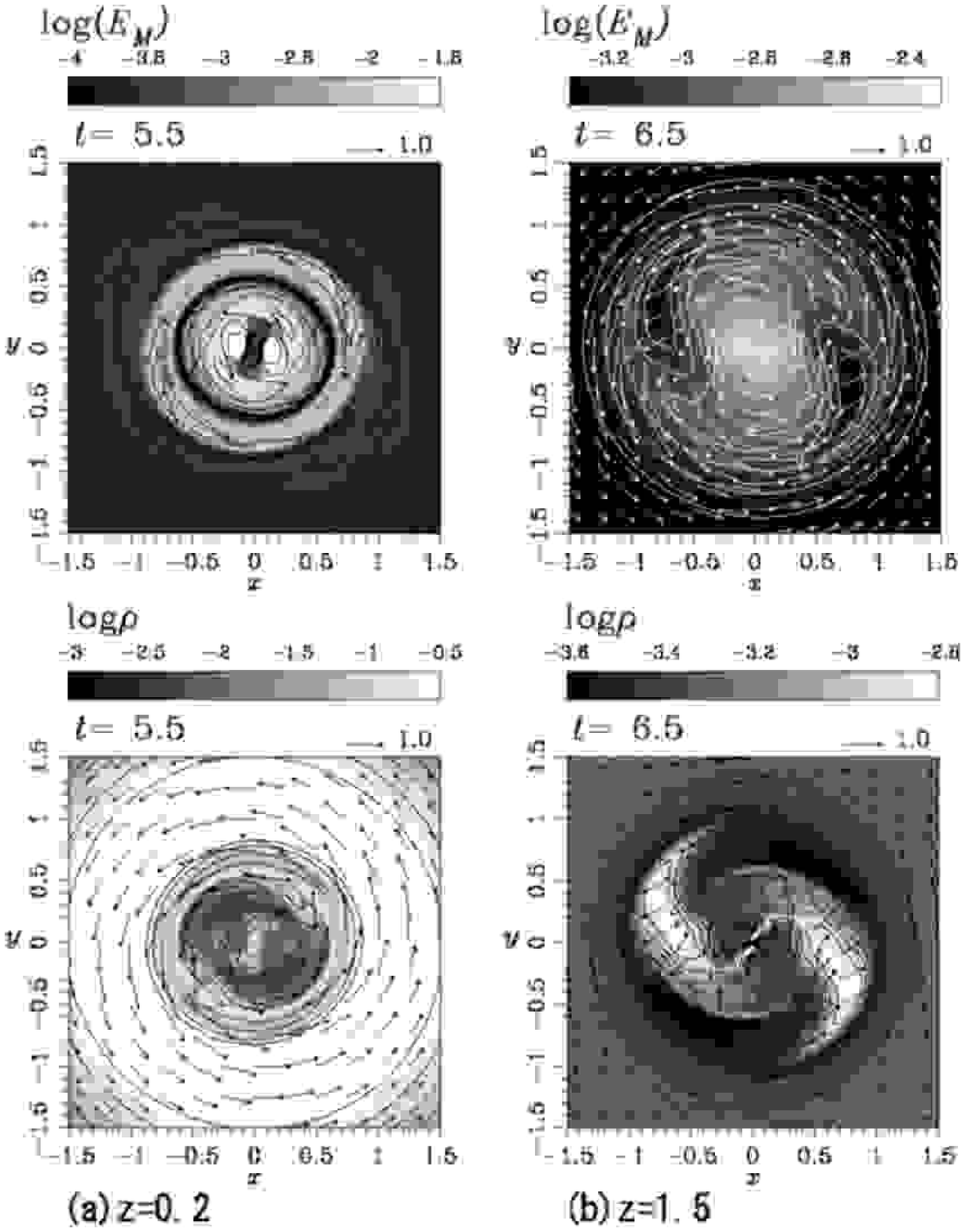}
\caption{The distribution of the magnetic energy and the logarithmic density 
(a)on the $z=0.2$ plane 
(in the disk) at $t=5.5$ and (b)on the $z=1.5$ plane (in the jet) at $t=6.5$ 
in the model S6.}
\label{FIG16}
\end{figure}

Figure \ref{FIG15} shows the time evolution of Fourier power spectra of the 
magnetic energy in the model S6. In the cases of the disk with sinusoidal 
perturbation ($\delta v_\phi \propto \sin 2 \phi$), the calculations have the 
periodicity of $\Delta \phi = \pm \pi$, so that we calculate the power spectra 
with even azimuthal wave number. The $m=2$ mode is dominant from the first 
stage of the run in the disk (see Figure \ref{FIG15}a) because the initial 
perturbation is of pure $m=2$ mode. 
The $m=2$ mode spectrum becomes almost constant level, and then increases 
continuously with oscillation. The $m=4$ and $6$ 
modes also start to grow at almost the same time with the $m=2$ mode. Their 
growth is also with oscillation and with almost the same growth rate. Fourier 
power spectra of the magnetic energy in the jet (see Figure \ref{FIG15}b) are 
not calculated before $t=5.4$ because the jet is not launched before that. 
The $m=2$ mode is dominant in the jet as well as in the disk. 
The power spectra of all the non-axisymmetric modes become nearly constant 
with time in the jet.
The each mode spectrum in the jet is not 
with oscillation by contrast with the growth of each mode in the disk. 
Figure \ref{FIG16} shows the distribution of the magnetic energy 
and the logarithmic density on 
the $z=0.2$ plane (in the disk) at $t=5.5$ and on the $z=1.5$ plane (in the 
jet) at $t=6.5$. The non-axisymmetric structure with $m=2$ is clearly seen. 

\begin{figure}
\figurenum{17}
\epsscale{1.0}
\plotone{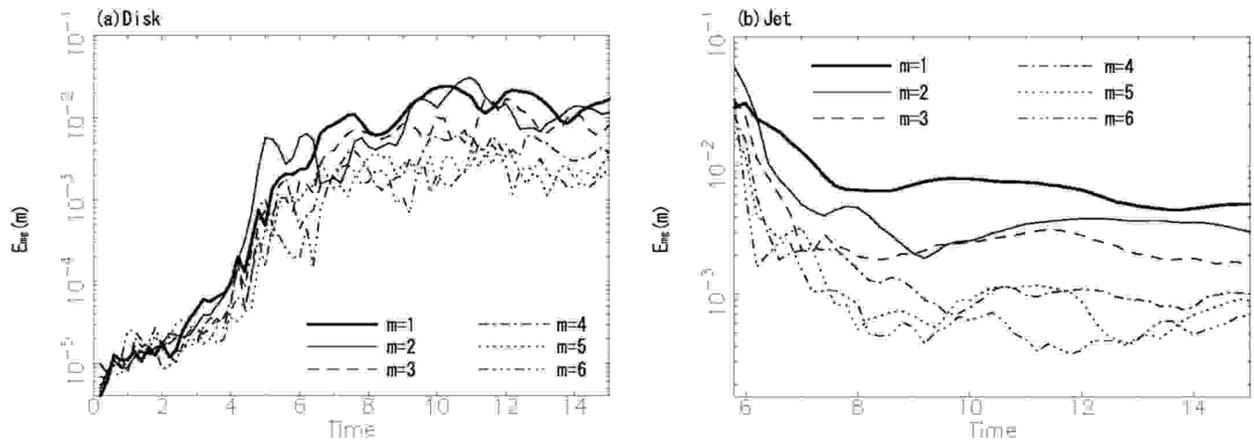}
\caption{Similar to Fig. \ref{FIG15}, but for the model R6.}
\label{FIG17}
\end{figure}

\begin{figure}
\figurenum{18}
\epsscale{0.9}
\plotone{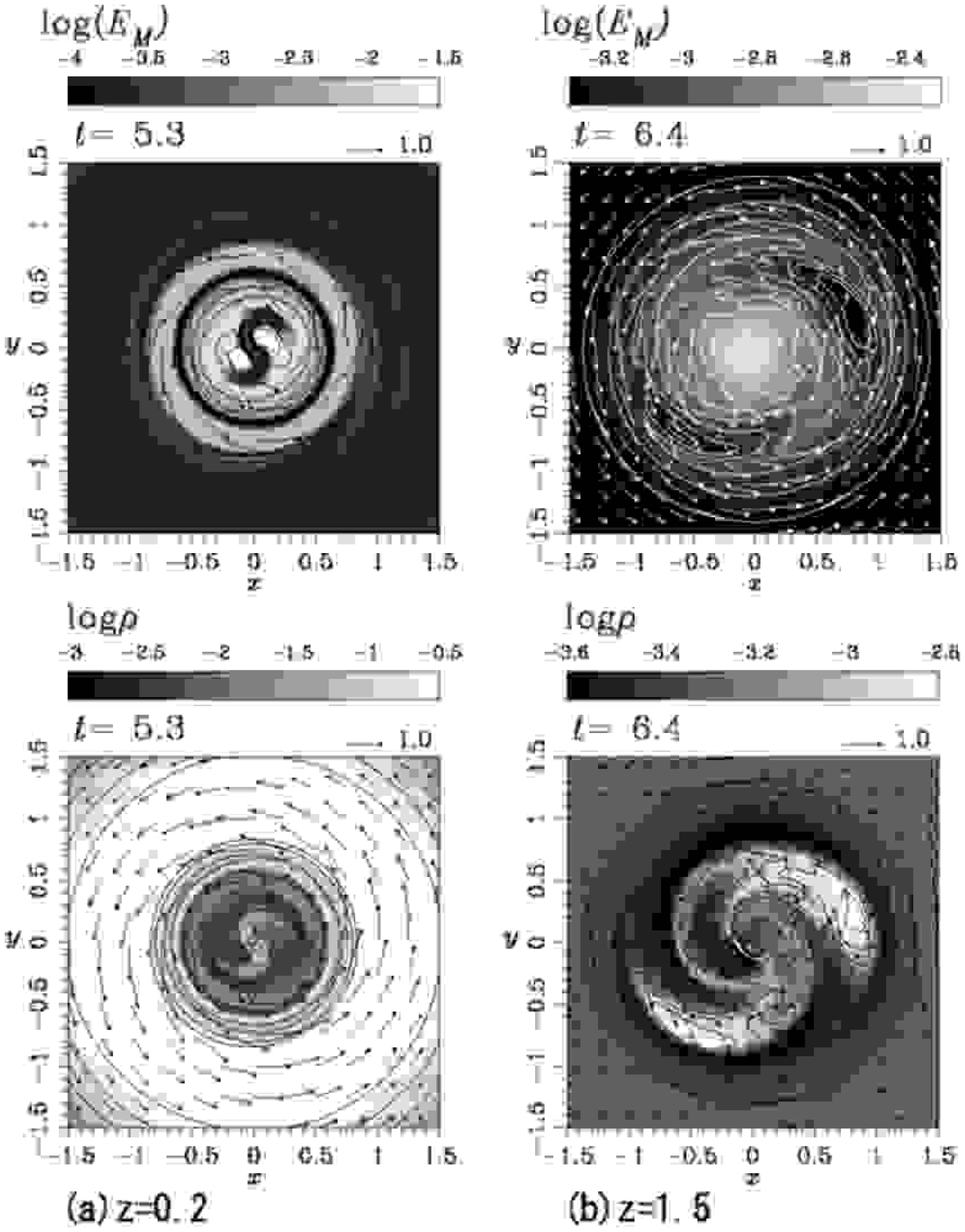}
\caption{The distribution of the magnetic energy and the logarithmic density 
(a)on the $z=0.2$ plane 
(in the disk) at $t=5.3$ and (b)on the $z=1.5$ plane (in the jet) at $t=6.4$ 
in the model R6.}
\label{FIG18}
\end{figure}

Figure \ref{FIG17} shows the Fourier power spectra of the magnetic energy in 
the model R6 as functions of time. In this case the initial perturbation has 
no particular mode, 
so that all modes show the growth with almost the same growth rate in the 
first stage in the disk (see Figure \ref{FIG17}a). 
The $m=2$ mode becomes dominant around $t=5.0$ in the disk. The $m=2$ mode 
spectrum, however, decreases for a time. At the same time, the growth rates of 
other modes also decrease. These can be thought to be related to the jet 
ejection. After the jet ejection, the $m=2$ mode becomes 
comparable to or predominant over the $m=1$ mode though, roughly speaking, 
the Fourier power spectrum is larger the longer (the smaller) the azimuthal 
wave length (wave number) is. 
As same as in the model S6, the growth of each mode in the disk shows 
oscillation though it can not be seen in the jet. It can 
be thought that the oscillation in the disk is related to the mass accretion 
(see also the bottom panel of Figure \ref{FIG13}). 
The $m=2$ mode is initially dominant in the jet, which is probably connected 
to the peak in the disk at $t=5.0$. In response to the decrease of the $m=2$ 
mode spectrum in the disk, that in the jet also decreases. However the $m=2$ 
mode spectrum becomes comparable to the $m=1$ mode spectrum later. It is to be 
noted that the power spectra of all the non-axisymmetric modes become nearly 
constant with time in the jet as well as in the case of the sinusoidally 
perturbed disk.
Figure \ref{FIG18} shows the distribution of the magnetic energy and the 
logarithmic density on the $z=0.2$ plane (in the disk) at $t=5.3$ and on the 
$z=1.5$ plane (in the jet) at $t=6.4$.

\subsection{Origin of the Non-axisymmetric Structure}
What mechanism makes the non-axisymmetric structure? 
In the previous section, it is made clear that the non-axisymmetric structure 
in the jet originates in the accretion disk. In this section, we 
discuss the cause of the non-axisymmetric structure in the disk. At first, 
the possibility of the magnetorotational instability (MRI; Balbus \& Hawley 
1991) is examined. The dispersion relation for the non-axisymmetric 
disturbance was given by Balbus \& Hawley (1992) (see the equation 2.24 in 
that paper). The dispersion relation equation is numerically solved; from the 
simulation result (e.g., Figure \ref{FIG18}a) $\lambda_r = 0.4$. The radial 
coordinate, 
$R$ in the equation, is assumed to be 1.0, where the avalanche-like accretion 
flow is initially remarkably formed. The axial wave length $\lambda_z$ is 
assumed to be 0.35, 
which is almost equal to the most unstable wavelength, $\lambda \sim 2 \pi 
V_A/\Omega$. From the initial 
condition, $\bm{V}_A = (V_{Ar},V_{A\phi},V_{Az}) = (0,0,0.056)$ and 
$\Omega = 1.0$ at $r=1.0$. The epicyclic frequency $\kappa$ is zero because of 
the constant angular momentum disk. The azimuthal wave number $m$ is the 
parameter.

One of the four roots is the unstable solution in all the ($m$=1$\sim$6) 
cases. The growth rate is higher the smaller the azimuthal wave number $m$ 
is, although the $m=2$ growth rate is comparable to the $m=1$ growth rate. 
The calculated growth rate, $\omega$, of the $m=2$ mode is about 0.54, and 
$\rm{exp} \left( \omega t \right)$ at $t=3.0$ is 5.1. 
The numerical result shows the increase of the $m=2$ mode spectrum by a 
factor of 5.9. These are consistent with the initial evolution of the modes 
in the disk (see Figure \ref{FIG17}a). 
After about $t=3.0$, the mode spectra show the non-linear growth. The temporal 
dominance of $m=2$ mode in the disk in the model R6 is at the MRI non-linear 
growth stage.

We briefly comment on the possibility of other instabilities. As showed 
in \S4.1, the K-H instabilities are stable between the jet and the ambient 
gas. How about between the accretion disk and the corona? In our simulations, 
they seem to be stable or the growth rates seem to be small 
because no growth of the K-H instabilities is seen in the hydrodynamic 
(no magnetic field) simulation. 
Another probable instability is a magnetic analogue of the non-axisymmetric 
hydrodynamical instability in the constant angular momentum torus, so-called 
Papaloizou-Pringle instability (Papaloizou \& Pringle 1984; Goldreich, 
Goodman, \& Narayan 1986; Hawley 1990). Curry \& Pudritz (1996) 
investigated the growth rate of the instability in various conditions. 
According to their result, the (magnetic) PP instability seems not to make 
an important contribution for forming the non-axisymmetric structure. 
The magnetic field is relatively weak (normalized Alfv\'en velocity, $V_z$ 
in Curry \& Pudritz, is equal to 0.056) so that it is almost hydrodynamic. 
Indeed, the plasma-$\beta$ in the disk is of the order of a hundred. In such
case, the growth rate is small.


\subsection{Amplification of Magnetic Field in the Disk}\label{AmpMag}
\begin{figure}
\figurenum{19}
\epsscale{1.0}
\plotone{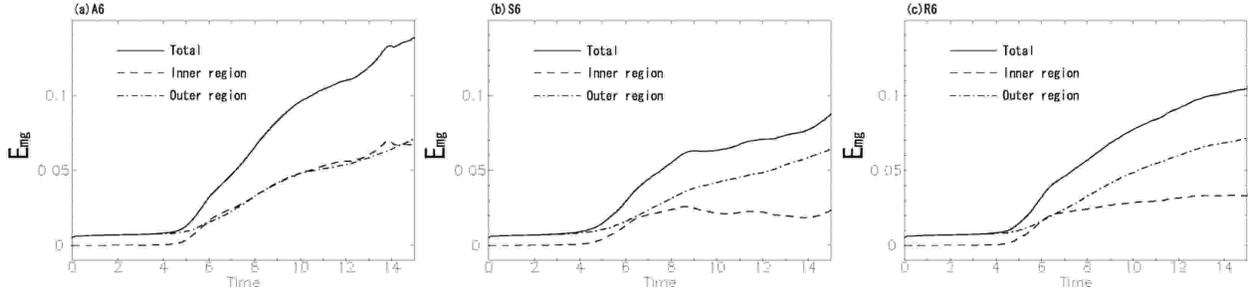}
\caption{Time evolution of the space-integrated magnetic energy in the 
accretion disk in each perturbation case. The solid line shows the magnetic 
energy in all the region of the disk. The broken or dash-dotted 
line shows the magnetic energy only in the inner ($r \le 0.6$) or outer 
($r > 0.6$) region of the disk respectively.}
\label{FIG19}
\end{figure}

Differentially rotating disks with a magnetic field amplify its magnetic field 
by the MRI. In this 
section we investigate the difference of the amplification depending on the 
form of the perturbation. Figure \ref{FIG19} shows the time variation of the 
magnetic energy in the disk. The definition of the disk is the same as that in 
the previous section. The solid line is the magnetic energy in all the region 
of the disk ($\int \bm{B}^2/8\pi \ dV_{disk}$), where $V_{disk}$ is the volume 
of the disk. The broken line is the magnetic energy only in the inner region 
($r \le 0.6$) of the disk and the dash-dotted line is the magnetic energy 
only in the outer region ($r > 0.6$) of the disk. The magnetic energy in the 
disk is most amplified in the axisymmetric case. The magnetic energy in the 
inner region displays a striking difference between the axisymmetric case and 
the other two cases though the magnetic energy in the outer region is almost 
the same. The total magnetic energy at $t=15.0$, therefore, shows the 
difference. The differences from the value for the model A6 are -0.051 and 
-0.034 for the model S6 and R6 respectively.

The time variation of the magnetic energy is described by the following 
equation.
\begin{equation}
\frac{\partial}{\partial t} \left( \frac{\bm{B}^2}{8\pi} \right) = 
- \bm{v} \cdot \left( \bm{J} \times \bm{B} \right) 
- \frac{1}{4\pi}\nabla \cdot \left( \bm{E} 
\times \bm{B}\right)\label{EQvamagene}
\end{equation}
where $\bm{J} = \left( \nabla \times \bm{B} \right)/4\pi$.
The first term in the right hand means the conversion of magnetic energy to 
kinetic energy 
by the Lorentz force. The second term is the transportation of the 
energy by Poynting flux. It is investigated below that a difference of 
the value of the first or second term in the equation (\ref{EQvamagene}) can 
explain the difference of the magnetic energy.

\begin{figure}
\figurenum{20}
\epsscale{1.0}
\plotone{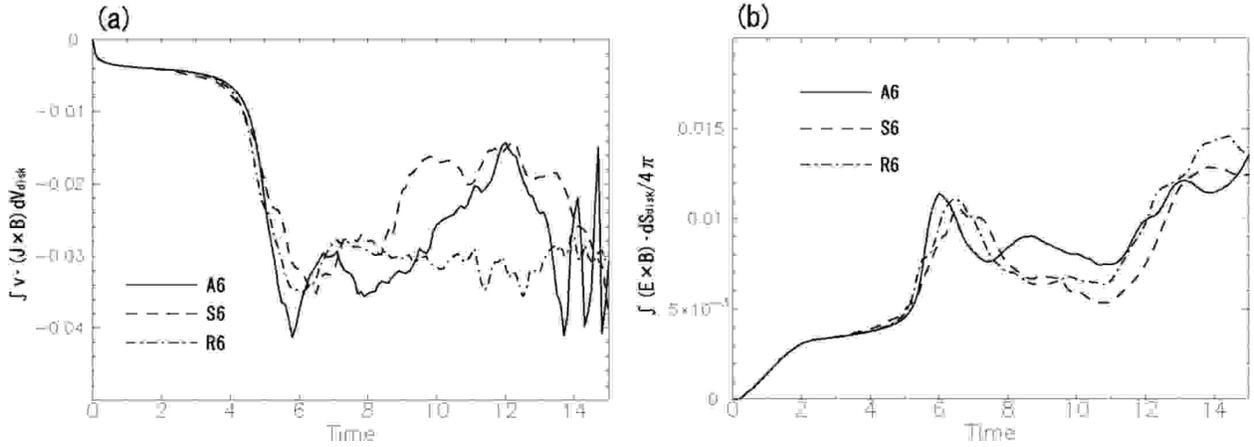}
\caption{(a)Time evolution of the work done by the Lorentz force in the 
accretion disk. A negative value means kinetic energy is converted to 
magnetic energy. (b)Time evolution of Poynting energy flowing out 
through a ``disk surface'' (see the text as regards the detailed 
definition of the ``disk surface'').}
\label{FIG20}
\end{figure}

Figure \ref{FIG20}a shows the work done by the Lorentz force in the disk 
($\int \bm{v} \cdot \left( \bm{J} \times \bm{B} \right) dV_{disk}$; a negative 
value means kinetic energy is converted to magnetic energy). 
The time integration of the work done by Lorentz force between $t=0.0$ and 
$t=15.0$ is done to investigate the origin of the difference in the magnetic 
energy between the different models, A6, S6, and R6.
For discussing the 
amplification of the magnetic energy, the sign should be inverted. 
The results are 
0.32, 0.26, and 0.34 for the model A6, S6, and R6 respectively. The 
differences from the value for the model A6 are -0.051 and 0.022 for the model 
S6 and R6 respectively. 

The another term is the energy transportation by Poynting flux. To calculate 
this value, the vector normal to the disk surface must be determined. 
However that is difficult, so that, for simplicity, the ``disk surface'' is 
defined as the $z=1.5$ plane ($r \le 2.0$) and the $r=2.0$ plane ($z \le 
1.5$). Figure \ref{FIG20}b shows the value of $\int \left( \bm{E} \times 
\bm{B} \right)/4\pi \ dS_{disk}$ as functions of time for the model A6 
(solid line), S6 (broken line), and R6 (dash-dotted line). The results of 
the time integral of this value between $t=0.0$ and $t=15.0$ are -0.11, -0.10, 
and -0.11 for the model A6, S6, and R6 respectively (the sign is inverted for 
discussing the amplification of the magnetic energy). The differences from 
the value for the model A6 are 0.01 (S6) and 0.0 (R6).

As regards the model S6, the difference of the magnetic energy from that for 
the model A6 at $t=15.0$ can 
be explained by the difference of the work done by the Lorentz force, 
ignoring the contribution of Poynting flux. On the other hand, the magnetic 
energy for the model R6 should become larger than that for the model A6 by 
0.022. The magnetic energy for the model R6 at $t=15.0$ is, however, smaller 
than that for the model A6 by 0.034. It means that the magnetic energy is 
dissipated by 0.056. The dissipation takes place at the inner region $(r \le 
0.6)$ of the disk (see Figure \ref{FIG19}). One possibility is the numerical 
magnetic reconnection. To confirm this point, resistive MHD simulations are
necessary. 
It is to be noted that the magnetic reconnection in the innermost region of 
the accretion disk is proposed to be the origin of X-ray shots observed in 
black hole candidates (Machida \& Matsumoto 2003). 
In any case, the physical process at the inner region of the 
disk seems to be important.


\subsection{Parameters in the Disk for Each Case}
Table \ref{ARS}, \ref{SPRS}, and \ref{RPRS} summarize several physical values 
calculated from the simulation result of the axisymmetric, the sinusoidal 
perturbation, and the random perturbation case respectively. 
The second column shows the initial space-averaged magnetic energy in the 
disk. 
The space-average of a certain physical quantity, $f$, is calculated as 
follows: $\langle f \rangle = \int_{V_s} f dV/V_s$, where 
$V_s$ is the volume of the disk. The definition of the disk 
is the same as in \S 4.2.
The third one shows the maximum space-averaged magnetic energy in the 
disk. The fourth one indicates the maximum amplification rate of the magnetic 
energy in the disk. The runs with the sinusoidal perturbation display the 
smallest amplification rates in all the magnetic field strength cases compared 
with the other two perturbation models. This 
is explained by the fact that in the sinusoidal perturbation case the work 
done by the Lorentz force is the smallest (see section \ref{AmpMag}). The rate 
for the random perturbation case is comparable to that for the axisymmetric 
case when the initial magnetic field strength is relatively small; however 
the rate for the random perturbation case becomes clearly smaller than that 
for the axisymmetric case as the initial magnetic field becomes strong. 
This may be because the magnetic energy dissipates by the numerical magnetic 
reconnection as discussed in section \ref{AmpMag}.

The fifth column shows the time-averaged ratio of the $r\phi$-component of 
the space-averaged Maxwell stress to the space-averaged 
pressure. This value corresponds to the non-dimensional viscous parameter 
(so-called $\alpha$ parameter). The $\alpha$ parameter means the efficiency of 
the angular momentum transport. In all the three 
(axisymmetric, sinusoidal perturbation, or random perturbation) cases, this 
parameter $(\equiv \alpha_{B,\phi R})$ increases with increasing the initial 
magnetic field strength. This is equivalent to the fact that the mass 
accretion rate increases with increasing the initial magnetic field. 
The sixth column is the time-averaged ratio of the ${\phi}z$-component of the 
space-averaged Maxwell stress to the space-averaged pressure. This value 
represents the efficiency of the angular momentum transport in the vertical 
direction. When we denote $\alpha_{B,\phi R} \propto E_{mg}^b$, we find $b$ 
is $0.66 \sim 0.75$; 
on the other hand $b$ becomes $0.49 \sim 0.54$ 
when $\alpha_{B,\phi Z}$ is fitted.
However, $\alpha_{B,\phi R}$ and $\alpha_{B,\phi Z}$ are comparable in the 
wide range of initial magnetic field strength. 
The last column shows the space- and time-averaged plasma-$\beta$ in the disk.

Finally we comment on the fact that the dynamical time of our simulations 
is about 2 orbital periods (strictly speaking, the dynamical time ranges from 
one a bit smaller than 2 orbital periods to about 2.5 orbital periods, 
depending on the initial magnetic field strength). 
In our simulations, the numerical instability often occurred at around 2 
rotations of the inner part of the disk by the violently deformed magnetic 
field
\footnote{We have not yet still fully understood the mechanism of the 
numerical instability, but it is related to many factors, including numerical 
methods, finite grids, physical parameters, etc. For example, MOC-CT scheme 
is known to cause the numerical instability when the current sheet is created 
(e.g., Hawley \& Stone 1995). If very low-$\beta$ plasma is created even 
locally, that region becomes the site of the numerical instability, since the 
magnetic energy is dominant so that any small error of the magnetic field 
affects the plasma dynamics significantly. Our problems are such ones that 
include locally low-$\beta$ plasma: even if the magnetic field is initially 
weak in a disk, it is significantly amplified by the differential rotation of 
the disk to produce locally low-$\beta$ regions.}.
It has often 
been said that the jets in nonsteady MHD simulations are transient phenomena 
owing to the particular choice of the initial conditions. However, Ibrahim, 
Sato, \& Shibata (2004), K. Sato et al. (in preparation), and A. Ibrahim et 
al. (in preparation) performed long-term simulations of MHD jet though the 
parameters of the accretion disk are different from the parameters in this 
paper. They showed 
that time-averages of the jet velocity, mass outflow rate, and mass accretion 
rate reproduce the scaling law found in Kudoh et al. (1998) although the jets 
are nonsteady and intermittent. These 
facts indicate the significance of our results regardless of the short 
calculation time. It is, of course, interesting whether the instabilities in 
the jet itself (e.g., the kink and/or K-H instabilities) develop after 
a long-time evolution. This is an important future work. 

\section{Conclusions}
We performed 3-dimensional MHD simulations of jet formation by solving the 
accretion disk self-consistently. The accretion disk is perturbed with the 
non-axisymmetric sinusoidal or random fluctuation of the rotational velocity 
to investigate the stability of MHD jet ejected from the disk in 3-dimension. 
Our results are summarized as follows:

\begin{enumerate}
\item The jet launched from the accretion disk whose rotational velocity 
is non-axisymmetrically perturbed has the properties that the jet velocity is 
proportional to $B_0^{1/3}$ and that the mass outflow rate is proportional 
to $B_0$, where $B_0$ is the initial magnetic field strength. 
The dependence of the mass accretion rate ($\dot{M}_a$) on the initial 
magnetic field strength is given by $\dot{M}_a \propto B_0^{1.4}$. 
The first relation is consistent with the scaling law of the Michel's 
solution (Michel 1969) and the result of the axisymmetric simulation (Kudoh 
et al. 1998). The other two relations are also consistent with Kudoh et al. 
(1998). Part of fluid elements, however, show the trajectories remarkably 
different from those in the axisymmetric case. This is caused by the decrease 
in the magnetic pressure gradient and centrifugal force in the radial 
direction those are against the magnetic tension (hoop stress).
\item The jet in both perturbation cases shows the non-axisymmetric 
structure with $m=2$, where $m$ means the azimuthal wave number. 
As the origin of the non-axisymmetric structure, the K-H 
instability is ruled out according to the stability condition. Calculation of 
the Fourier power spectra of the magnetic energy shows that the $m=2$ 
mode becomes dominant in the jet at least once. The time evolution of the 
power spectra indicates that the non-axisymmetric structure in the jet 
originates in the accretion disk, not in the jet itself. 
In both perturbation cases, all the non-axisymmetric modes in the jet 
reach almost constant levels after about 1.5 orbital periods of the accretion 
disk, while all modes in the accretion disk grow with oscillation. 
\item The magnetic energy is amplified in the accretion disk by the 
magnetorotational instability (MRI). In the non-axisymmetric cases, 
amplification rates are smaller than that in the axisymmetric case. 
The difference between the axisymmetric case and the non-axisymmetric cases 
is notable in the inner region of the accretion disk. In the sinusoidal 
perturbation case, this is explained by the difference of the work done 
by the Lorentz force. In the random perturbation case, however, that can not 
be explained by such difference or the difference of the energy transportation 
by Poynting flux. One possibility is the numerical reconnection in the inner 
region of the accretion disk.
\item The space- and time-averaged ratio of the Maxwell stress component 
that is related to the angular momentum transport in the radial direction to 
gas pressure, $\langle \langle -B_r B_\phi/4\pi \rangle / \langle p \rangle 
\rangle \equiv \alpha_{B,\phi R}$, shows the dependence on the initial 
magnetic field strength like $\alpha_{B,\phi R} \propto B_0^{1.4}$. That for 
the Maxwell stress component that is related to the angular momentum transport 
in the vertical direction shows the dependence like $\langle \langle -B_\phi 
B_z/4\pi \rangle / \langle p \rangle \rangle \equiv \alpha_{B,\phi Z} \propto 
B_0^{1.0}$. However, $\alpha_{B,\phi Z}$ is comparable to $\alpha_{B,\phi R}$ 
in the wide range of the initial magnetic field strength.
\end{enumerate}

This work is partly supported by the Japan Society for the Promotion of 
Science Japan-UK Cooperation Science Program (principal investigators: 
K. Shibata and N. O. Weiss), and the Grant-in-Aid for the 21st Century COE 
"Center for Diversity and Universality in Physics" from the Ministry of 
Education, Culture, Sports, Science and Technology (MEXT) of Japan.
Numerical computations were carried out on VPP5000 at the
Astronomical Data Analysis Center of the National Astronomical
Observatory, Japan (project ID: yhk32b and rhk05b), which is an
interuniversity research institute of astronomy operated by the Ministry
of Education, Culture, Sports, Science, and Technology.



\clearpage
\begin{table}
\caption{Normalization Units}\label{TABnorm}
\begin{tabular}{ccc}
\hline \hline
Physical Quantities & Description & Normalization Unit \\ \hline
$t$ & Time & $r_0/V_{K0}$\\
$r, z$ & Length & $r_0$\\
$\rho$ & Density & $\rho_0$\\
$p$ & Pressure & $\rho_0 V_{K0}^2$\\
$\bm{v}$ & Velocity & $V_{K0}$\\
$\bm{B}$ & Magnetic field & $( \rho_0 V_{K0}^2)^{1/2}$\\
\hline
\end{tabular}
\tablecomments{The unit length $r_0 = (L_0^2/GM)^{1/(1 - 2a)}$ is the radius 
of the density maximum in the initial disk. The unit velocity $V_{K0} 
\equiv (GM/r_0)^{1/2}$ is the Keplerian velocity at $(r, z) = (r_0, 0)$. The 
unit density $\rho_0$ is the initial density at $(r, z) = (r_0, 0)$. It is 
assumed that $a = 0$ throughout this paper}
\end{table}


\clearpage
\begin{table}
\caption{Models and Parameters}\label{TABMandP}
\begin{tabular}{|c|c|c|c|c|c|}
\hline
 & & & \multicolumn{3}{c|}{Perturbation}\\ \cline{4-6}
$E_{mg}$ & $\beta_{0d}$\tablenotemark{a} & $\beta_{0c}$\tablenotemark{b} & 
None & Sinusoidal & Random\\ \hline
$1.0 \times 10^{-5}$ & 10200 & 200 & A1 & S1 & R1\\
$2.0 \times 10^{-5}$ &  5100 & 100 & A2 & S2 & R2\\
$5.0 \times 10^{-5}$ &  2040 &  40 & A3 & S3 & R3\\
$1.0 \times 10^{-4}$ &  1020 &  20 & A4 & S4 & R4\\
$2.0 \times 10^{-4}$ &   510 &  10 & A5 & S5 & R5\\
$5.0 \times 10^{-4}$ &   204 &   4 & A6 & S6 & R6\\
$1.0 \times 10^{-3}$ &   102 &   2 & A7 & S7 & R7\\
$2.0 \times 10^{-3}$ &    51 &   1 & A8 & S8 & R8\\
\hline
\end{tabular}
\tablenotetext{a}{The value of plasma-$\beta$ ($= 8\pi p/\bm{B}^2$) at 
$(r,z)=(1,0)$ in the disk.}
\tablenotetext{b}{The value of plasma-$\beta$ at $(r,z)=(0,1)$ in the corona.}
\tablecomments{This table shows how to name the models. The number of the 
model name indicates 
the magnetic field strength and the alphabet shows the type of perturbation 
in the rotational velocity of the accretion disk (A: There is no perturbation. 
This means that initial condition is ``Axisymmetric'' . S; Sinusoidal 
perturbation. R; Random perturbation).}
\end{table}

\clearpage
\begin{table}
\begin{center}
\caption{Axisymmetric Runs}\label{ARS}
\begin{tabular}{ccccccc}
\hline \hline
Model & 
$\langle E_{Md0}\rangle$\tablenotemark{a,b} & 
$\langle E_{Md_{max}} \rangle$\tablenotemark{c} & 
$\left( (3)-(2) \right)/(2)$ & 
$\langle \langle -\frac{B_r B_{\phi}}{4\pi} \rangle / \langle p \rangle \rangle$\tablenotemark{d} & 
$\langle \langle -\frac{B_{\phi} B_z}{4\pi} \rangle / \langle p \rangle \rangle$ & 
$\langle \langle \beta_{d} \rangle \rangle$\\
(1) & (2) & (3) & (4) & (5) & (6) & (7)\\
\hline
A1 & $5.0 \times 10^{-6}$ & $1.6 \times 10^{-4}$ & 31 & 0.0035 & 0.0068 & $1.4 \times 10^{2}$\\
A2 & $1.0 \times 10^{-5}$ & $3.3 \times 10^{-4}$ & 32 & 0.0079 & 0.012 & 84\\
A3 & $2.5 \times 10^{-5}$ & $6.0 \times 10^{-4}$ & 23 & 0.018 & 0.022 & 41\\
A4 & $5.0 \times 10^{-5}$ & $1.0 \times 10^{-3}$ & 19 & 0.030 & 0.032 & 23\\
A5 & $1.0 \times 10^{-4}$ & $1.7 \times 10^{-3}$ & 16 & 0.051 & 0.044 & 13\\
A6 & $2.5 \times 10^{-4}$ & $3.2 \times 10^{-3}$ & 12 & 0.089 & 0.063 & 6.2\\
A7 & $5.0 \times 10^{-4}$ & $5.7 \times 10^{-3}$ & 10 & 0.14 & 0.082 & 3.4\\
A8 & $1.0 \times 10^{-3}$ & $7.7 \times 10^{-3}$ & 6.7 & 0.17 & 0.093 & 2.0\\
\hline
\end{tabular}
\tablenotetext{a}{$\langle \ \rangle$ denotes the space average.}
\tablenotetext{b}{$E_{Md0}$ is the magnetic energy in the disk ($z \le 1.5$ and $\Theta \ne 0.0$) at $t=0.0$.}
\tablenotetext{c}{$E_{Md_{max}}$ is the maximum value of the magnetic energy in the disk between $t=0.0$ and $t=11.6$.}
\tablenotetext{d}{$\langle \langle \ \rangle \rangle$ denotes the space and 
time average in the disk.}
\end{center}
\end{table}

\clearpage
\begin{table}
\begin{center}
\caption{Sinusoidal Perturbation Runs}\label{SPRS}
\begin{tabular}{ccccccc}
\hline \hline
Model & 
$\langle E_{Md0}\rangle$\tablenotemark{a,b} & 
$\langle E_{Md_{max}} \rangle$\tablenotemark{c} & 
$\left( (3)-(2) \right)/(2)$ & 
$\langle \langle -\frac{B_r B_{\phi}}{4\pi} \rangle / \langle p \rangle \rangle$\tablenotemark{d} &
$\langle \langle -\frac{B_{\phi} B_z}{4\pi} \rangle / \langle p \rangle \rangle$ &
$\langle \langle \beta_{d} \rangle \rangle$\\
(1) & (2) & (3) & (4) & (5) & (6) & (7)\\
\hline
S1 & $5.0 \times 10^{-6}$ & $1.2 \times 10^{-4}$ & 22 & 0.0027 & 0.0061 & $1.4 \times 10^{2}$\\
S2 & $1.0 \times 10^{-5}$ & $2.1 \times 10^{-4}$ & 20 & 0.0061 & 0.011 & 87\\
S3 & $2.5 \times 10^{-5}$ & $4.5 \times 10^{-4}$ & 17 & 0.014 & 0.021 & 44\\
S4 & $5.0 \times 10^{-5}$ & $7.1 \times 10^{-4}$ & 13 & 0.024 & 0.032 & 26\\
S5 & $1.0 \times 10^{-4}$ & $1.1 \times 10^{-3}$ & 9.5 & 0.041 & 0.047 & 14\\
S6 & $2.5 \times 10^{-4}$ & $2.0 \times 10^{-3}$ & 7.0 & 0.073 & 0.066 & 6.7\\
S7 & $5.0 \times 10^{-4}$ & $3.4 \times 10^{-3}$ & 5.8 & 0.11 & 0.083 & 3.8\\
S8 & $1.0 \times 10^{-3}$ & $4.7 \times 10^{-3}$ & 3.7 & 0.15 & 0.10 & 2.2\\
\hline
\end{tabular}
\tablenotetext{a}{$\langle \ \rangle$ denotes the space average.}
\tablenotetext{b}{$E_{Md0}$ is the magnetic energy in the disk ($z \le 1.5$ and $\Theta \ne 0.0$) at $t=0.0$.}
\tablenotetext{c}{$E_{Md_{max}}$ is the maximum value of the magnetic energy in the disk between $t=0.0$ and $t=10.7$.}
\tablenotetext{d}{$\langle \langle \ \rangle \rangle$ denotes the space and 
time average in the disk.}
\end{center}
\end{table}

\clearpage
\begin{table}
\begin{center}
\caption{Random Perturbation Runs}\label{RPRS}
\begin{tabular}{ccccccc}
\hline \hline
Model & 
$\langle E_{Md0}\rangle$\tablenotemark{a,b} & 
$\langle E_{Md_{max}} \rangle$\tablenotemark{c} & 
$\left( (3)-(2) \right)/(2)$ & 
$\langle \langle -\frac{B_r B_{\phi}}{4\pi} \rangle / \langle p \rangle \rangle$\tablenotemark{d} &
$\langle \langle -\frac{B_{\phi} B_z}{4\pi} \rangle / \langle p \rangle \rangle$ &
$\langle \langle \beta_{d} \rangle \rangle$\\
(1) & (2) & (3) & (4) & (5) & (6) & (7)\\
\hline
R1 & $5.0 \times 10^{-6}$ & $1.7 \times 10^{-4}$ & 33 & 0.0041 & 0.0056 & $1.5 \times 10^{2}$\\
R2 & $1.0 \times 10^{-5}$ & $3.4 \times 10^{-4}$ & 33 & 0.0090 & 0.010 & 88\\
R3 & $2.5 \times 10^{-5}$ & $7.1 \times 10^{-4}$ & 27 & 0.021 & 0.020 & 44\\
R4 & $5.0 \times 10^{-5}$ & $9.8 \times 10^{-4}$ & 19 & 0.033 & 0.031 & 26\\
R5 & $1.0 \times 10^{-4}$ & $1.4 \times 10^{-3}$ & 13 & 0.048 & 0.045 & 15\\
R6 & $2.5 \times 10^{-4}$ & $2.6 \times 10^{-3}$ & 9.3 & 0.079 & 0.067 & 6.9\\
R7 & $5.0 \times 10^{-4}$ & $3.9 \times 10^{-3}$ & 6.9 & 0.12 & 0.084 & 3.9\\
R8 & $1.0 \times 10^{-3}$ & $5.6 \times 10^{-3}$ & 4.6 & 0.14 & 0.10 & 2.3\\
\hline
\end{tabular}
\tablenotetext{a}{$\langle \ \rangle$ denotes the space average.}
\tablenotetext{b}{$E_{Md0}$ is the magnetic energy in the disk ($z \le 1.5$ and $\Theta \ne 0.0$) at $t=0.0$.}
\tablenotetext{c}{$E_{Md_{max}}$ is the maximum value of the magnetic energy in the disk between $t=0.0$ and $t=10.2$.}
\tablenotetext{d}{$\langle \langle \ \rangle \rangle$ denotes the space and 
time average in the disk.}
\end{center}
\end{table}

\clearpage

\clearpage

\clearpage

\clearpage

\clearpage

\clearpage

\clearpage

\clearpage

\clearpage

\clearpage

\clearpage

\clearpage

\clearpage

\clearpage

\clearpage

\clearpage

\clearpage

\clearpage

\clearpage

\end{document}